\documentclass[floatfix,aps,prb,twocolumn,superscriptaddress,showpacs]{revtex4-1}
\usepackage{amsmath,amssymb,graphicx}

\def\ea{{\it et al.}}

\def\dds{dd\sigma} \def\ddp{dd\pi} \def\ddd{dd\delta}
\def\sds{sd\sigma} \def\sss{ss\sigma}
 
\def\e{{\rm e}} \def\E{\varepsilon}
\def\sh{\frac{\scriptstyle 1}{\scriptstyle 2}}
\def\Edis{E_{\text{dis}}} \def\Eads{E_{\text{ads}}}
\def\Ebind{E_{\text{bind}}}
\def\Ebindv{E^v_{\text{bind}}(n)}
\def\half{\genfrac{}{}{}{1}{1}{2}}

\begin{document}

\title{Electronic structure and total energy of interstitial hydrogen
       in iron: Tight binding models}

\author{A. T. Paxton}
\email{Tony.Paxton@iwm.fraunhofer.de}
\affiliation{Fraunhofer Institut f\"ur
  Werkstoffmechanik IWM, W\"ohlerstr.~11, 79108 Freiburg,
  Germany}
\affiliation{Karlsruher Institut f\"ur Technologie, Institut
  f\"ur Zuverl\"assigkeit von Bauteilen und Systemen (IZBS),
  Kaiserstr.~12, 76131 Karlsruhe, Germany}
\affiliation{Atomistic Simulation Centre, School of Mathematics
  and Physics, Queen's University Belfast, Belfast BT7 1NN, UK}
\author{C. Els{\"a}sser}
\email{Christian.Elsaesser@iwm.fraunhofer.de}
\affiliation{Fraunhofer Institut f\"ur
  Werkstoffmechanik IWM, W\"ohlerstr.~11, 79108 Freiburg,
  Germany}
\affiliation{Karlsruher Institut f\"ur Technologie, Institut
  f\"ur Zuverl\"assigkeit von Bauteilen und Systemen (IZBS),
  Kaiserstr.~12, 76131 Karlsruhe, Germany}

\begin{abstract}
An application of the tight binding approximation is presented
for the description of electronic structure and interatomic force
in magnetic iron, both pure and containing hydrogen
impurities. We assess the simple canonical $d$-band description in
comparison to a non orthogonal model including $s$~and $d$~bands.
The transferability of our models is tested against
known properties including the segregation energies of hydrogen
to vacancies and to surfaces of iron. In many cases agreement is
remarkably good, opening up the way to quantum mechanical
atomistic simulation of the effects of hydrogen on mechanical
properties. 
\end{abstract}


\maketitle

\section{Introduction}
\label{sec_Intro}

In this paper we demonstrate tight binding (TB) models for iron
with and without interstitial hydrogen impurites at the
concentrated and dilute limits. Although there is a large number
of existing classical potentials which are certainly of great
importance and usefulness, they all suffer from a particular
drawback in that the underlying classical EAM-type models for
pure Fe, with one apparent exception,\cite{EAC09} fail to predict
the known core structures of screw
dislocations.\cite{EAC09,Mrovec09} On the other hand tight
binding models abstracted into bond order potentials correctly predict
core structures in agreement with first principles
calculations.\cite{Mrovec09} Ultimately one of the many
goals is to study how interstitials form atmospheres around
dislocations and impede or enhance flow through mechanisms such
as hydrogen enhanced local plasticity\cite{Lynch88} (HELP) and so
it is essential that dislocation core structures are correctly
predicted. A further slightly disturbing feature of the classical
models is the truly vast number of parameters involved which have
to be fitted to a very large training set of data. Here in common
with the approach to classical model fitting,\cite{EAC09} we
first construct models for pure iron and then go on to make
models for hydrides without further adjusting the Fe--Fe
interaction parameters. But in contrast we try to keep the number
of parameters and fitting targets to a minimum and focus on the
ability of the models to {\it predict} those properties that are
normally included in the training sets in the construction of
classical potentials. We would argue that this is possible
because the TB approximation comprises a correct quantum
mechanical description of both magnetism and the metallic and
covalent bond and so the correct physics is built in from the
start. That being the case, we do not expect the theory to be
over sensitive to the choice of parameters and indeed in the
procedure we describe below a large number of equally useful
models is thrown up.

The structure of this paper is as follows. In
section~\ref{sec_tba} we revisit the tight binding approximation
and discuss its parameters and their environment dependence, or
{\it screening}. We describe two models for pure~Fe in
section~\ref{sec_PureFe} which are fitted to properties of bulk
bcc $\alpha$-Fe and used to predict properties of fcc $\gamma$-Fe
and hcp $\epsilon$-Fe, as well as surface and vacancy formation
energies in $\alpha$-Fe. In section~\ref{sec_FeH} we
augment one of these models with Fe--H interactions which we fit
to the properties of four monohydride FeH phases, and test
against adiabatic potential surfaces. We then proceed
to the dilute limit of~H in~Fe in
section~\ref{sec_predictions-FeH} without further adjustment of
parameters and use our model to predict segregation energies of~H
to interstitial sites, vacancies and surfaces of $\alpha$-Fe. By
and large, we find remarkable agreement with published
experimental results and {\it ab initio} calculations. We discuss
our models and conclude in section~\ref{sec_discussion}.

\begin{figure*}
\caption{\label{fig_Fe-bands} (color online) Energy bands for
  bcc Fe at its experimental lattice constant, 2.87~\AA. The
  coloring is such that $s$ character is green and $d$ character
  is blue. The Fermi energy is indicated by a horizontal
  line. The upper panels are majority and the
  lower panels minority spin bands. Far left is the tight binding
  $d$-band model and in the center our non orthogonal~$sd$
  model. To the right are bands calculated in the LSDA-GGA.}
\begin{center}
\includegraphics[scale=1.1]{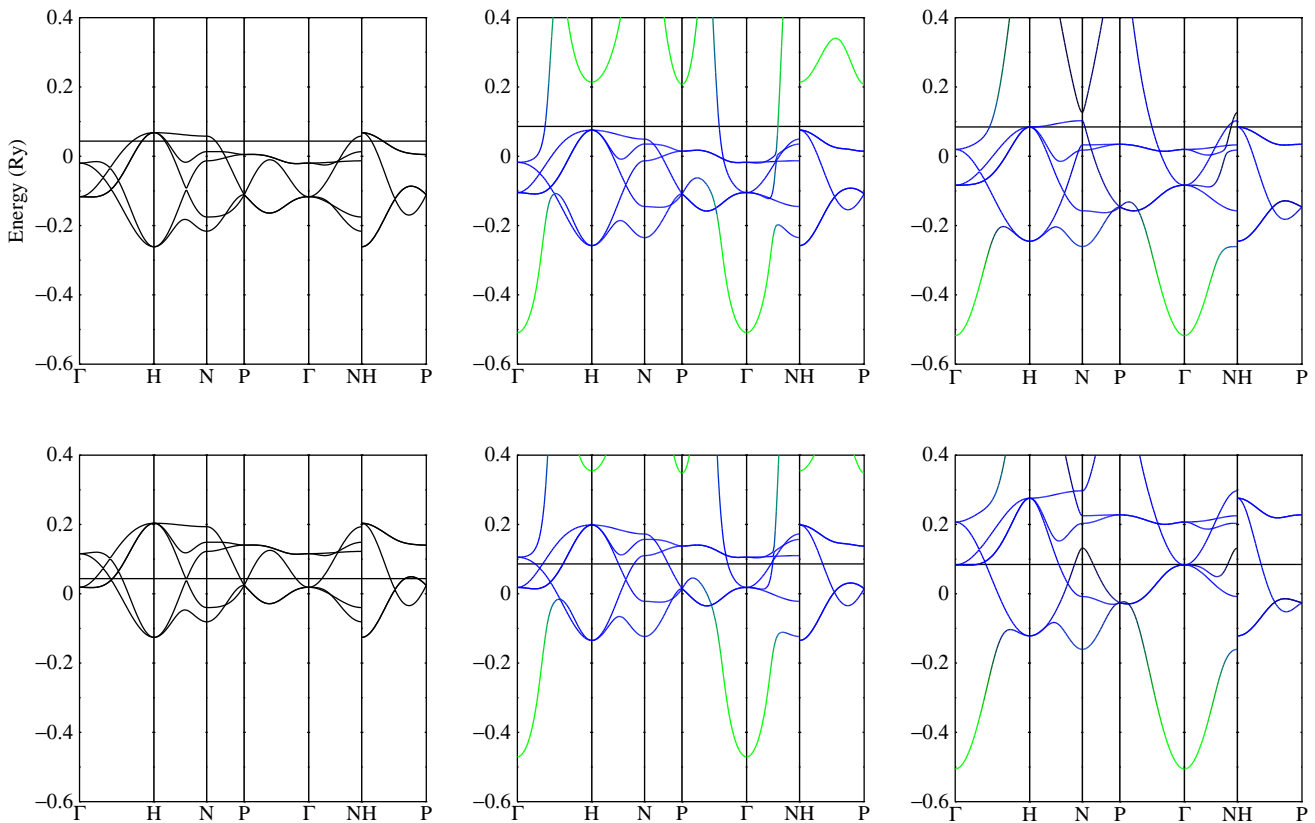}
\end{center}
\end{figure*}

\section{The tight binding approximation and transferability}
\label{sec_tba}

\subsection{Distance scaling and range of the hopping integrals}
\label{subsec_scaling}

There is no need to rehearse the tight binding approximation in
any detail here. Recently Paxton and Finnis\cite{Paxton08}
constructed tight binding models for magnetic Fe and Fe--Cr
alloys and details can be found there as well as in many other
publications.\cite{Harrison80,Pettifor95,Finnis03,Liu05,Paxton09}
However we do wish to make some preliminary remarks. The scheme
that we use is the self consistent Stoner model for itinerant
ferromagnetism\cite{Liu05} and goes beyond the fixed moment and
rigid band approximations. The connection between tight binding
theory and the first principles local spin density approximation
(LSDA) to density functional theory (DFT) is now well
established.\cite{Foulkes89,Sutton88,Finnis03} TB is
computationally several orders of magnitude faster than LSDA
because the hamiltonian is constructed from a look-up table of
parameterized hopping integrals, $h$, and possibly overlap
integrals, $s$. These are conventionally written in
Slater and Koster's notation\cite{Slater54} as $\sss$, $\sds$,
$\dds$, $\ddp$, $\ddd$. Central to a tight binding model is the
way in which these integrals scale with bond length. In this work
we will use\cite{Ducastelle70,Spanjaard84,Paxton96,Paxton08}
\begin{equation}
\label{eq_hop}
    h(r) = h_0\e^{-qr}
\end{equation}
and similarly for overlap integrals, when used,
\begin{equation}
\label{eq_s}
    s(r) = s_0\e^{-qr}.
\end{equation}
The alternative is to use the power law scaling, $h\sim r^{-n}$,
demanded by canonical band
theory.\cite{Heine67,Andersen73,Andersen75,Pettifor77} There is
no strong argument to prefer one over the other; in fact by
equating\cite{Skinner91} logarithmic derivates of $h(r)$ at, say
first neighbors at a distance $r_0$, we have $n=qr_0$ and in the
bcc structure of Fe $q\approx 1$~a.u. corresponds to the canonical
$n=5$ (see table~\ref{tbl_parms-PureFe}, below).

This brings us to a well known paradox of tight binding modeling
namely that the decay of the hopping integrals is known {\it a
  priori} from band theory, which may render them longer ranged
than is desirable. A well known example is the group~IV
semiconductors where by analogy with the free electron bands, to
reproduce the volume dependence of the bandwidth the hopping
integrals must scale with $n=2$.\cite{Froyen79} This scaling is
bound to lead to very long ranged hopping integrals; on the other
hand it is known that the first neighbor approximation is the
right one, and attempts to include further neighbors
fail.\cite{Paxton87} For many purposes it is adequate simply to
cut off the interactions between first and second neighbors, but
this can lead to difficulties in work on complex defects or
in molecular dynamics. An elegant solution was provided by
Goodwin, Skinner, and Pettifor\cite{Goodwin89} (GSP) which cuts
off a power law exponentially beyond some chosen cut-off
distance, $r_c$. There are two drawbacks to this. ({\it i\/})~An
exponential decay can still lead to discontinuities in molecular
dynamics (as one still needs to impose a cut-off in the neighbor
lists).  ({\it ii\/})~The GSP form maintains the {\it value} but
not the {\it slope} of the underlying power law at first
neighbors. Therefore our preference is to retain the power or
exponential scaling given by the canonical band theory and to
choose two distances, $r_1$ and $r_c$, between which to smoothly
augment the interaction to zero. This can be achieved by matching
value, slope and curvature at $r_1$ and at $r_c$ with
a fifth degree polynomial which replaces the hopping integral in
that range.\cite{Znam03} We show our hopping integrals thus
augmented at figure~\ref{fig_hopping} in section~\ref{sec_FeH}
below, where we discuss this matter further.

\subsection{The pair potential, transferability, and non orthogonality}
\label{subsec_trans}

The hopping integrals provide an attractive force, which in the
conventional tight binding models is balanced by a repulsive pair
potential, which here may take the form
\begin{equation}
\label{eq_pp}
     \phi(r) = B_1 \e^{-p_1r} - B_2 \e^{-p_2r}
\end{equation}
in which, as suggested by Liu~\ea\cite{Liu05}, both $B_1$ and
$B_2$ are positive. This potential is expected to be repulsive at
short range but is not positive for all $r$ (see
Fig.~\ref{fig_pairFe}, below).

An additional non pairwise repulsion is provided if it is chosen
to make the model basis non orthogonal. This may give a number of
advantages.\cite{Paxton08} One is, that it is widely believed
that non orthogonality confers a greater transferability to the
model.\cite{Allen86} By this is meant that a model constructed
for a particular crystal structure is less likely to fail when
transferred into a situation of different crystal structure or
increased or reduced coordination. We will wish to focus
critically on this aspect of our models below. It is instructive
at this stage to recall that by its very construction the tight
binding approximation discards all three center terms in the
hamiltonian.\cite{Paxton09} On the one hand the canonical band
theory shows that these, like non orthogonality, are of second
order in the band width.\cite{Pettifor72,Pettifor77,Andersen84}
On the other hand Tang~\ea\cite{Tang96} and Haas~\ea\cite{Haas98}
took the important step of proposing {\it environment dependent}
hopping integrals. In this {\it empirical} scheme the hopping
integral between two atoms is modified in the close proximity of
a third atom---in the extreme limit this third atom may approach
the two center bond, generally speaking weakening or
``screening'' it, and eventually come in between the two
atoms. Whereas the screening was first described by an empirical
formula, Pettifor succeeded in {\it deriving} the
Tang~\ea\cite{Tang96} expression from the L\"owdin transformation
of the non orthogonal hamiltonian.\cite{Pettifor00} In particular
he showed that $sd$ overlap matrix elements in pure transition
metals provide this ``screening'' of the two center
bond. Therefore rather than adopting explicit environment
dependence as is done in recent bond order
potentials,\cite{Mrovec04,Mrovec07} we retain the two center
approximation and employ non orthogonal models to account for the
screening.

\subsection{The choice of parameters}
\label{subsec_choice}

A related and highly significant finding\cite{Pettifor00} is that
hopping integrals extracted from an LSDA hamiltonian calculated
using the tight binding LMTO-ASA method\cite{Andersen84} are
discontinuous between first and second neighbors in bcc
transition metals. These discontinuities are described
consequently by the screening---a feature of the geometry of the
bcc lattice---leading to the analytic form of
Tang~\ea\cite{Tang96} and Haas~\ea\cite{Haas98} The point we wish
to raise here is that it became clear\cite{Pettifor00} that
transferable hopping integrals may be extracted from an LSDA
hamiltonian thus avoiding the usual need for
fitting.\cite{Pettifor00,Paxton09} Of course there is no unique
tight binding model for a given element since the LSDA
hamiltonian is basis-set dependent. We do not adopt this approach
here for two reasons. First, the hopping integrals deduced from
the LMTO-ASA\cite{Pettifor00,Paxton09} derive from a hamiltonian
whose {\it on-site} matrix elements are strongly volume dependent
whereas in the tight binding approximation these terms are volume
independent and hence any volume dependence of the electronic
structure must be taken up by the scaling
law~(\ref{eq_hop}). Second, if the hopping integrals and their
scaling are taken from {\it ab initio} bandstructures without
permitting further adjustment, then essential properties such as
elastic constants, lattice constants and structural energy
differences may have to rely on the choice of pair potential
placing a large burden on that part of the model which is the
most {\it ad hoc}.

\section{Models of pure iron}
\label{sec_PureFe}

\begin{center}
\begin{table}
\caption{\label{tbl_parms-PureFe} Parameters of our tight binding
  models for pure Fe. The \{$h$\} and \{$s$\} are the $h_0$ and
  $s_0$ of equations~(\ref{eq_hop}) and~(\ref{eq_s}). All
  quantities are given in atomic Rydberg units (1~bohr =
  0.529~\AA, 1~Ry = 13.61~eV). Note that in the minimal basis,
  orthogonal~$d$ model, the number of $d$-electrons, $N_d$ is a
  parameter.\cite{Liu05} Both hopping integrals and pair
  potentials are smoothly cut off between distances $r_1$ and
  $r_c$. These are shown in units of the bcc lattice constant,
  $a$. Both pair potentials are cut off with $r_1=1.1$ and
  $r_c=1.4$, that is, between second and third neighbors of the
  bcc lattice (see Fig.~\ref{fig_pairFe} below). By expressing
  $r_1$ and $r_c$ in units of $a$ we imply that these scale with
  the lattice constant, for example in the calculation of the
  bulk modulus, so that in a perfect lattice first and second
  neighbors always see the ``proper'' pair
  potential~(\ref{eq_pp}) and hopping integrals. This also
  applies below (Fig.~\ref{fig_FeH-EV}) to energy--volume curves
  in FeH, both for first and second bcc neighbors and first fcc
  neighbours.}
\begin{tabular}{|c|c|c|c|c|c|c|c|c|}
\hline
model              & \multicolumn{4}{c|}{orthogonal~$d$} &      \multicolumn{4}{c|}{non orthogonal~$sd$}  \\
\hline
  $\E_s-\E_d$   & \multicolumn{4}{c|}{---}    &   \multicolumn{4}{c|}{0.15}    \\
\hline
 & $h_0$/$s_0$ & $q$ & $r_1$ & $r_c$ & $h_0$/$s_0$ & $q$ & $r_1$ & $r_c$ \\
\hline
  $h_{\dds}$      & --4.464  & 1       & 1.1 & 1.4 &   --2.438    & 0.9  & 1.1 & 1.4    \\   
  $h_{\ddp}$      &   2.976  & 1       & 1.1 & 1.4 &     1.997    & 0.9  & 1.1 & 1.4   \\   
  $h_{\ddd}$      & --0.744  & 0.94    & 1.1 & 1.4 &   --0.907    & 0.9  & 1.1 & 1.4    \\   
  $h_{sd}$        &   ---    &  ---    & --- & --- &   --0.141    & 0.3  & 1.1 & 2.0   \\   
  $h_{ss}$        &   ---    &  ---    & --- & --- &   --0.350    & 0.3  & 1.1 & 2.0   \\   
  $s_{sd}$        &   ---    &  ---    & --- & --- &     0.50     & 0.6  & 1.1 & 2.0   \\
  $s_{ss}$        &   ---    &  ---    & --- & --- &     0.45     & 0.5  & 1.1 & 2.0   \\ 
\hline
  $N_d$           & \multicolumn{4}{c|}{6.80}    &          \multicolumn{4}{c|}{---}    \\
  $I$             & \multicolumn{4}{c|}{0.050}   &          \multicolumn{4}{c|}{0.055}    \\   
\hline
  $B_1$           & \multicolumn{4}{c|}{1248.0}  &          \multicolumn{4}{c|}{536.0}    \\   
  $p_1$           & \multicolumn{4}{c|}{1.4510}  &          \multicolumn{4}{c|}{1.4900}   \\   
  $B_2$           & \multicolumn{4}{c|}{1025.0}  &          \multicolumn{4}{c|}{371.2}   \\   
  $p_2$           & \multicolumn{4}{c|}{1.4087}  &          \multicolumn{4}{c|}{1.4131}  \\   
\hline
\end{tabular}
\end{table}
\end{center}

\subsection{Orthogonal $d$, and non orthogonal $sd$ models}
\label{subsec_d-and-sd}

\begin{figure}
\caption{\label{fig_Fe-dos} Densities of states of pure bcc~Fe in
  the orthogonal~$d$ (a) and non orthogonal~$sd$ (b) tight
  binding models compared to the LSDA-GGA (c). Majority and
  minority spin states are shown in the upper and lower panels of
  each. The zero of energy is shifted to the Fermi energy.}
\begin{center}
\includegraphics[scale=0.4]{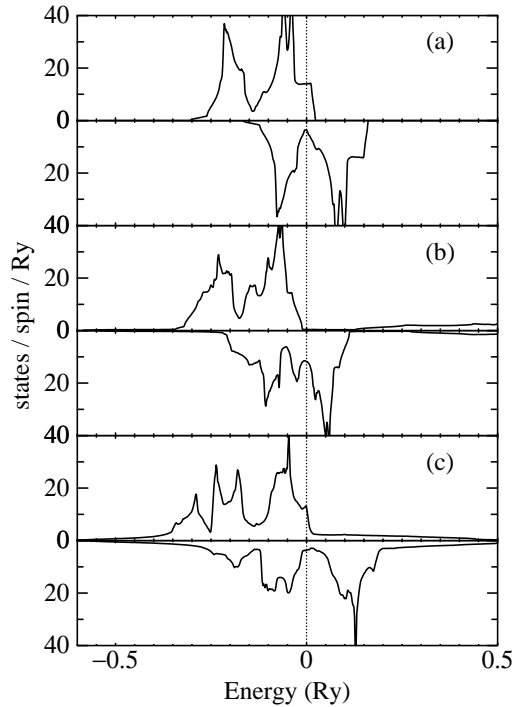}
\end{center}
\end{figure}

Construction of a tight binding model for transition metals is
quite straight forward if it is not required to take the
parameters from first principles bandstructure
calculations.\cite{Paxton08} Given that the scaling should be
close to canonical, as should the ratios,\cite{Pettifor77}
$$
    \dds\hskip 6pt :\hskip 6pt\ddp \hskip 6pt :\hskip 6pt \ddd \hskip 6pt 
    \approx -6 \hskip 6pt :\hskip 6pt 4 \hskip 6pt :\hskip 6pt -1
$$
it is simple enough to guess a set of hamiltonian and overlap
matrix elements and adjust these until the resulting energy bands
match reasonably closely those from the LSDA. In fact, in all
that follows we have used the LSDA with a generalized gradient
correction (GGA) of Perdew~\ea\cite{PBE} With the exception of
data taken from the literature all our LSDA-GGA results are
calculated using the full potential LMTO
method.\cite{Methfessel00} Energy bands calculated in this way
are shown on the right in Fig.~\ref{fig_Fe-bands}. A simple
canonical $d$-band model produces the bands shown to the left of
Fig.~\ref{fig_Fe-bands}. In addition to the integrals already
discussed, we require a Stoner parameter, $I$, which represents
an on-site Coulomb integral,\cite{Pettifor95,Paxton08} to achieve
a splitting of the up and down spins.\cite{Liu05,Paxton08}
Furthermore since the canonical model omits the $s$-band which is
occupied by roughly one electron\cite{Pettifor77} it is necessary
to fix a number of $d$~electrons, $N_d$.\cite{Paxton96,Liu05}
This is both the most simple and most reliable model for
transition
metals.\cite{Pettifor79,Williams80,Pettifor87,Pettifor95}
Nevertheless for the present purposes we wish to extend this
model. In the interests of transferability and to account for the
bond screening without explicit environment dependent bond
integrals, we explore here the addition of an $s$-orbital to the
basis, including $\sds$ and $\sss$ non orthogonality. We will
also give arguments for this necessity when we come to the iron
hydrides below. The resulting bands, again obtained by simple
comparison by eye with the LSDA-GGA bands are shown in
Fig.~\ref{fig_Fe-bands}. Densities of states associated with the
LSDA-GGA and tight binding models are shown in
Fig.~\ref{fig_Fe-dos}.

\begin{center}
\begin{table}
\caption{\label{tbl_props-PureFe} Calculated properties using
  parameters from table~\ref{tbl_parms-PureFe}. They are compared
  in the right hand column to either experimental values or
  values calculated using LSDA-GGA, the latter written in
  italics. A proper comparison of the cohesive energy,
  $E_{{\text{coh}}}$, with experiment should take account of the
  spin polarization energy of the free atom which is absent in
  the tight binding limit of infinite separation; this energy is
  as much as\cite{Philipsen96} 0.32~Ry so that the calculated
  $E_{{\text{coh}}}$ should amount to 0.63~Ry. Hence the apparent
  better agreement of the orthogonal~$d$ model is
  misleading. Both ferromagnetic (FM) and non magnetic (NM)
  fcc~Fe is included; we compare the experimental data to the FM
  calculations: the lattice constant is extrapolated to
  0${^{\circ}}$K;\cite{Seki05} the elastic constants are taken
  from phonon dispersion curves\cite{Zarestky87} measured at
  1428${^{\circ}}$K which is above the Curie temperature
  (1043${^{\circ}}$K) although local moments are expected to
  persist.\cite{Edwards83} LSDA-GGA~NM values in parentheses
  refer to the low spin phase.}
\begin{tabular}{|cc|c|c|c|}
\hline
 & & $d$ &  $sd$  & \\
\hline
 bcc & $a$ (\AA)           & 2.87  & 2.87  & 2.87, {\it 2.84}  \\
 bcc & $K$  (GPa)          & 175   & 184   & 170,\footnotemark[1] {\it 173} \\
 bcc & $C'$ (GPa)          & 48    & 43    & 52,\footnotemark[1] {\it 62} \\
 bcc & $c_{44}$ (GPa)      & 118   & 108   & 121,\footnotemark[1] {\it 109}  \\
 bcc & moment ($\mu_B$)    & 2.7   & 2.2   & 2.2  \\
 bcc & $E_{{\text{coh}}}$ (Ry) & 0.36  & 0.51  & 0.31  \\
 \hline
 hcp & $a$ (\AA)           & 2.54  & 2.51   & {\it 2.54}  \\ 
 hcp & $K$  (GPa)          & 164   & 171   & {\it 160}  \\
 hcp & moment ($\mu_B$)    & 2.4   & 1.8   & {\it 2.4} \\
 hcp &  $E_{mag}$ (mRy)     & 7.7   & 4.6   & {\it 7.7} \\
 \hline
 \multicolumn{2}{|c|}{$\Delta E_{{\text{coh}}}$ hcp--bcc (mRy)} & 12 & 3 & {\it 15}  \\
 \hline
 fcc  (FM)& $a$ (\AA)      & 3.68 & 3.60   & 3.55,\footnotemark[2] {\it 3.64} \\
 fcc  (FM)& $K$  (GPa)     & 223  & 187    & 133,\footnotemark[3] {\it 191} \\
 fcc  (FM)& $C'$ (GPa)     & 13   & 12     & 16,\footnotemark[3] {\it --88}    \\
 fcc  (FM)& $c_{44}$ (GPa) & 79   & 74     & 77,\footnotemark[3] {\it 13}\\
 fcc  (NM)& $a$ (\AA)      & 3.45 & 3.51 & {\it 3.46} {\it (3.45)} \\
 fcc  (NM)& $K$  (GPa)     & 358  & 232  & {\it 294}  {\it (294)}  \\
 fcc  (NM)& $C'$ (GPa)     & 96   & 72 &   {\it 102}  {\it (102)}\\
 fcc  (NM)& $c_{44}$ (GPa) & 227  & 151 &  {\it 250}  {\it (249)}\\
\hline
\end{tabular}
\footnotetext[1]{from data extrapolated from $3^{\circ}$K to
  $0^{\circ}$K by Adams~\ea\cite{Adams06}}
\footnotetext[2]{Reference [\onlinecite{Seki05}]}
\footnotetext[3]{Reference [\onlinecite{Zarestky87}]}
\end{table}
\end{center}

Having obtained two sets of bond integrals, we proceed to find
parameters of the pair potential and we do this by adjusting the
four parameters in~(\ref{eq_pp}) to the lattice constant and the
three elastic constants of bcc~Fe. This cannot be done exactly
because of the restricted form of the pair potential. The
parameter sets are given in table~\ref{tbl_parms-PureFe} and
resulting properties are in good agreement with experiment or
LSDA-GGA calculations as can be seen in
table~\ref{tbl_props-PureFe}.
\nocite{Philipsen96,Seki05,Zarestky87,Edwards83,Adams06}
Calculations, written in italics in table~\ref{tbl_props-PureFe},
have been done using the full potential LMTO
method\cite{Methfessel00} and elastic constants are all obtained
at the theoretical lattice constants. Our calculated lattice and
elastic constants are in general agreement with previous
work.\cite{Guo00}

\subsection{Predictions of the models}
\label{subsec_PureFe-predictions}

\subsubsection{Magnetic moment and structural magnetic energy differences}
\label{subsubsec_PureFe-struc}

\begin{figure*}
\caption{\label{fig_TB-Fe-EV} (color online) Contributions to
  the energy--volume relation in the orthogonal~$d$ (left) and
  non orthogonal~$sd$ (right) tight binding models. The lower
  panel shows the volume dependence of the magnetic moment. The
  dotted lines refer to the hcp crystal structure and show the
  rapid collapse of the moment under pressure. The solid line is
  the bcc moment and may be compared with the circles which are
  LSDA-GGA calculations. The upper panel shows the bandstructure
  energy (blue) and the magnetic energy (green) and their sum in
  red.  Solid lines refer to bcc and dotted lines to hcp. The
  pair potential energy favors bcc in both models. A vertical
  line indicates the equilibrium volume in bcc~Fe.}
\begin{center}
\includegraphics[scale=0.75]{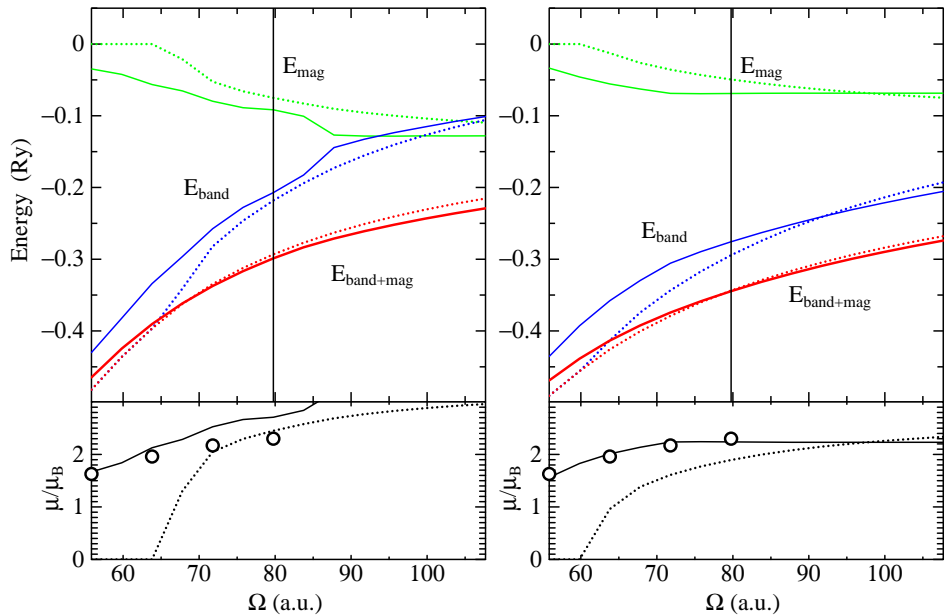}
\end{center}
\end{figure*}

\begin{figure}
\caption{\label{fig_pairFe} (color online) Pair potentials in the
  orthogonal~$d$ (dotted line) and non orthogonal~$sd$ (solid
  line) tight binding models. Note how these are negative at some
  of the critical bond lengths. While this helps to stabilize bcc
  against hcp, the real benefit is in obtaining a correct elastic
  constant $C'$. For reference below we also show here the Fe--H
  pair potential in blue (see
  table~\ref{tbl_parms-FeH}). Vertical lines are placed at first
  and second neighbor distances in bcc~Fe and at first neighbor
  distances in hcp~Fe.}
\begin{center}
\includegraphics[scale=0.5]{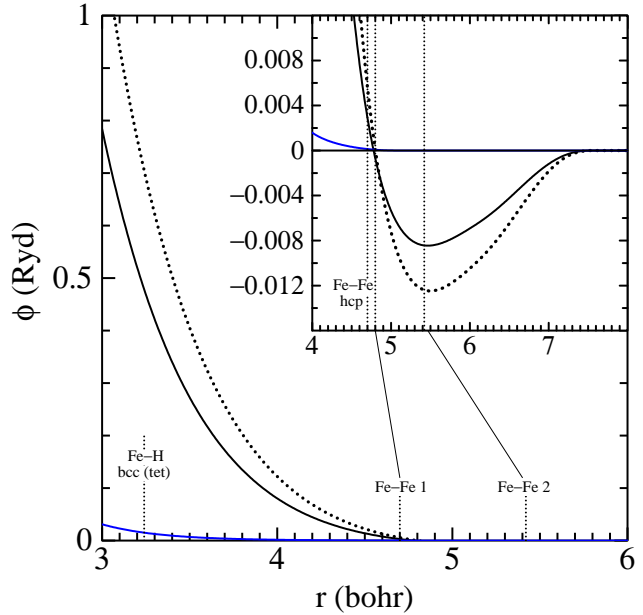}
\end{center}
\end{figure}

It should be noted that the two models we have described are
rather intuitively obtained and so, apart from the pair potential
it cannot be said that these are ``fitted'' in the sense of a
classical potential. Hence the properties shown in
table~\ref{tbl_props-PureFe} are in essence {\it predictions} of
the model, validating the underlying correctness of the tight
binding theory. These predictions can be discussed in more detail
by reference to Fig.~\ref{fig_TB-Fe-EV} which shows the structural
energy--volume relation in bcc and hcp~Fe broken down into
bandstructure energy and magnetic energy
contributions.\cite{Paxton08} Both models reproduce the essential
features which are, ({\it i\/}) the rapid collapse of the hcp
magnetic moment under pressure; ({\it ii\/}) the slow decline of
the bcc moment and ({\it iii\/}) the stabilization of bcc over
hcp being a result of the magnetism. The role of the pair
potential warrants explanation here. Fig.~\ref{fig_pairFe} shows
equation~(\ref{eq_pp}) plotted using the parameters of our two
models from table~\ref{tbl_parms-PureFe}. It might well be
supposed that the stability of the bcc phase compared to the hcp
is an artefact of the negative region of $\phi(r)$ falling at the
second neighbors of the bcc structure, while the 12 hcp nearest
neighbor distances fall in a positive region. This would be a
valid criticism of our and Liu~\ea's\cite{Liu05} models but is
misleading. In fact we find that we can easily make models that
stabilize bcc employing a pair potential that is positive
everywhere. In addition the stabilisation of the bcc structure
can be amplified by choosing larger Stoner $I$ parameter. We
allow a larger moment in our orthogonal~$d$ model since it is
known that the magnetic moment in bcc~Fe would be closer to
$2.6\mu_{B}$ in the absence of $sd$ and $pd$
hybridization.\cite{Heine80,Pettifor83} Therefore the LSDA-GGA
bcc--hcp energy difference is better rendered in that
model (table~\ref{tbl_props-PureFe}) whereas we have chosen a
value of $I$ in our non orthogonal~$sd$ model that strikes a
compromise between a smaller bcc--hcp energy difference having
the benefit of a magnetic moment closer to the observed
value. The real benefit of the form~(\ref{eq_pp}) is that it
enables a sufficiently large value of the elastic constant $C'$
which otherwise appears too small. It is well known that $C'$ can
become very soft in bcc metals and the values we obtain in
table~\ref{tbl_props-PureFe} are the best we can achieve after
many trials with the other parameters and scalings in the
models. Indeed in the model of Liu~\ea\cite{Liu05}, $C'$ is
significantly lower than ours. The only solution we know of to
fit the elastic constants exactly is to employ a spline form for
the pair potential as is done in the fitting of bond order
potentials,\cite{Znam03,Mrovec04,Mrovec07} and we are rather
reluctant to make such a departure from physical intuition.

\subsubsection{fcc $\gamma$-Fe}
\label{subsubsec_PureFe-gamma}

Because our models were fitted to the bcc Fe lattice and elastic
constants, it is important to focus on the lower part of
table~\ref{tbl_props-PureFe} which deals with the fcc phase
of~Fe. This is $\gamma$-Fe which is the base for the austenitic
steels and the crystal structure adopted by pure~Fe above
1185${^{\circ}}$K.\cite{Leslie} It is well
known\cite{Kaufman63,Roy77,Christensen88,CE1} that $\gamma$-Fe
exists in a high spin ferromagnetic and a low spin (approximately
non magnetic) modification and we show predictions for both
phases in table~\ref{tbl_props-PureFe} which we compare with
LSDA-GGA calculations and experimental observations. It is a mark
of transferability that both models give a good account of each
of the two fcc phases. Neither model fully captures the large and
negative $C'$ or the softening of $c_{44}$ of the LSDA-GGA in
the high spin phase; although they are in better accord with
experiment than the LSDA-GGA, the proper comparison is with the
0${^{\circ}}$K calculations. The elastic softening in $\gamma$-Fe
is consistent with the measured temperature dependence of $C'$ in
the Invar alloys,\cite{Tajima76} therefore it is encouraging that
our models are able to describe this important physical
phenomenon at least in principle. It has already been shown that
elastic and phonon softening with {\it increasing} temperature in
$\alpha$-Fe is captured in the tight binding
approximation.\cite{Hasegawa85,Hasegawa87}

\begin{center}
\begin{table}
\caption{\label{tbl_surfaces-PureFe} Calculated surface
  energies in J/m$^2$. Values in parentheses are for truncated
  bulk (unrelaxed) surfaces. LSDA-GGA calculations are taken from
  Spencer~{\it et al.}\cite{Spencer02}}
\begin{tabular}{|c|c|c|c|}
\hline
model & orthogonal~$d$ & non orthogonal~$sd$  & GGA \\
\hline
(110) & 1.77 (1.77) & 1.53 (1.56) & 2.27 (2.27) \\
(001) & 2.12 (2.15) & 1.74 (1.79) & 2.29 (2.32) \\
(111) & 3.54 (3.85) & 2.80 (3.34) & 2.52 (2.62) \\
\hline
\end{tabular}
\end{table}
\end{center}

\subsubsection{Surface energies}
\label{subsubsec_PureFe-surfaces}

The proper test of transferability is to carry the models into
situations of over or under coordination. Here, we do this by
addressing the surface energies of pure Fe. We have set up the
(110), (001) and (111) surfaces of bcc~Fe and relaxed the atom
positions by energy minimization using the Hellmann--Feynman
forces.\cite{Finnis98b,Liu05,Paxton09} The resulting energies are
shown in table~\ref{tbl_surfaces-PureFe} in order of decreasing
coordination, the most close packed surface being (110). We
achieve modest, but satisfactory agreement with published
LSDA-GGA calculations\cite{Spencer02} at least for the two most
close packed surfaces. It is in fact notable that the LSDA-GGA
predicts all the surfaces to have nearly the same energy with
(111) being a little higher. This is not reflected in the tight
binding models, indicating limits to their transferability. The
orthogonal $d$~model gives the greater spread in energies,
demonstrating to some extent the greater transferability afforded
by the inclusion of an $s$~orbital. It is gratifying that both
models give a qualitative account of surface energies without
having been fitted, at least in the case of the (110) and (001)
the latter being of most importance as it's the usual cleavage
face.\cite{Allen56,Leslie,Ayer06} It is also significant in the
present context that the effect of H~on pure Fe and Fe--Si is to
enable cleavage also on the (110) planes.\cite{Nakasato78}

\subsubsection{Vacancy formation energy}
\label{subsubsec_PureFe-vacancies}
\nocite{Gillan89}

\begin{center}
\begin{table}
\caption{\label{tbl_PureFe-vacancies} Vacancy formation energy,
  $E_v^f$, in eV, of pure Fe, calculated with the orthogonal~$d$ and
  non orthogonal~$sd$ tight binding models and compared to published LSDA-GGA and
  experimental results.}
\begin{tabular}{|c|c|c|c|c|}
\hline
model & $d$ & $sd$  & LSDA-GGA & expt. \\
\hline
relaxed    & 2.39 & 1.33 & 1.95,\footnotemark[1] 2.18\footnotemark[2] & 1.61--1.75,\footnotemark[4] 1.59\footnotemark[5] \\
           &      &      & 2.09\footnotemark[3]                       & $2.0\pm 0.2$\footnotemark[6] \\
unrelaxed  & 2.42 & 1.36 & 2.24,\footnotemark[1] 2.60\footnotemark[2] &  \\
\hline
\end{tabular}
\footnotetext[1]{Reference [\onlinecite{Domain01}]}
\footnotetext[2]{Reference [\onlinecite{Soderlind00}]}
\footnotetext[3]{Reference [\onlinecite{Tateyama03}]}
\footnotetext[4]{Muon spin rotation,\cite{Furderer87,Seeger98}}
\footnotetext[5]{Quenching-in and electrical resisitivity\cite{Seydel94,Seeger98}}
\footnotetext[6]{Positron annihilation,\cite{DeSchepper83} but
  Seeger\cite{Seeger98} asserts that $E_v^f\lessapprox 1.85$~eV}
\end{table}
\end{center}

A further test of the transferability is to predict the formation
energy of a vacancy. We do this by constructing 54~and 53~atom
``supercells'' of bcc Fe ($3\times 3\times 3$ cubic two-atom unit
cells), one of which has an atom missing. The structure is
relaxed by energy minimization; its resulting total energy is
denoted $E(\text{Fe}_{53})$. The energy of the 54~atom supercell
is denoted $E(\text{Fe}_{54})$. Then the vacancy formation
energy, neglecting volume relaxation, is\cite{Gillan89}
\begin{equation*} 
E_v^f = E(\text{Fe}_{53}) - \frac{53}{54}\>E(\text{Fe}_{54}).
\end{equation*} 
Our results are shown in
table~\ref{tbl_PureFe-vacancies}\nocite{Domain01,Soderlind00,Furderer87,Seeger98,Seydel94}
which also gives values for the ``unrelaxed'' vacancy.  As for
the surface energies, $E_v^f$ is underestimated by the non
orthogonal~$sd$ model and overestimated by the orthogonal~$d$
model. The likely error compared to experiment in the latter
however is more than twice that of the non orthogonal~$sd$ model,
again demonstrating some benefit in transferability of including
the non orthogonal $s$-orbitals. 

\section{Adding F\lowercase{e}--H interactions}
\label{sec_FeH}

\begin{figure}
\caption{\label{fig_Interstitials} (color online) To
  illustrate the tetrahedral (T) and octahedral (O) interstices
  in the bcc (upper figure) and fcc (lower figure) crystals. Note
  that in the bcc lattice the octahedral site is at the center of
  a distorted octahedron, unlike the fcc where it is regular. The
  distance to the two apical atoms, shown here as a horizontal
  bond, is shorter by a factor $1/\sqrt{2}$ than the distances to
  the equatorial atoms, two of which are shown here in the upper
  face. This leads in general to the well known tetragonal
  distortion of the bcc lattice near octahedral interstitial
  atoms, for example in martensite. For details see
  refs~[\onlinecite{Barrett,Leslie}]. Neither is the tetrahedral
  interstitial site in the bcc lattice regular---indeed both
  octahedral and tetrahedral bcc interstices have tetragonal
  symmetry. The fcc crystal structure with all the octahedral
  sites occupied becomes that of cubic rocksalt adopted by many
  transition metal carbides and nitrides. In fcc, the tetrahedral
  site is regular; when half these sites are occupied the
  resulting crystal structure is that of zincblende.}
\begin{center}
\includegraphics[scale=0.4]{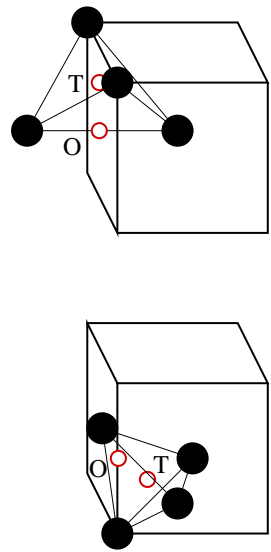}
\end{center}
\end{figure}

\begin{figure*}
\caption{\label{fig_FeH-EV} (color online) Cohesive energy and
  magnetic moment as a function of volume per Fe~atom in the four
  FeH phases calculated within the LSDA-GGA (left). Dotted lines
  denote non magnetic phases. The cohesive energy is with respect to
  solid $\alpha$-Fe and molecular H$_2$ also calculated using the
  same energy functional and hence is an approximation to the
  heat of formation. Note that on this basis none of the phases
  is expected to exist. On the right we show the same quantities
  calculated in the non orthogonal~$sd$ tight binding model. We
  expect that the almost exact degeneracy of bcc TET and fcc TET
  is accidental.}
\begin{center}
\includegraphics[scale=1]{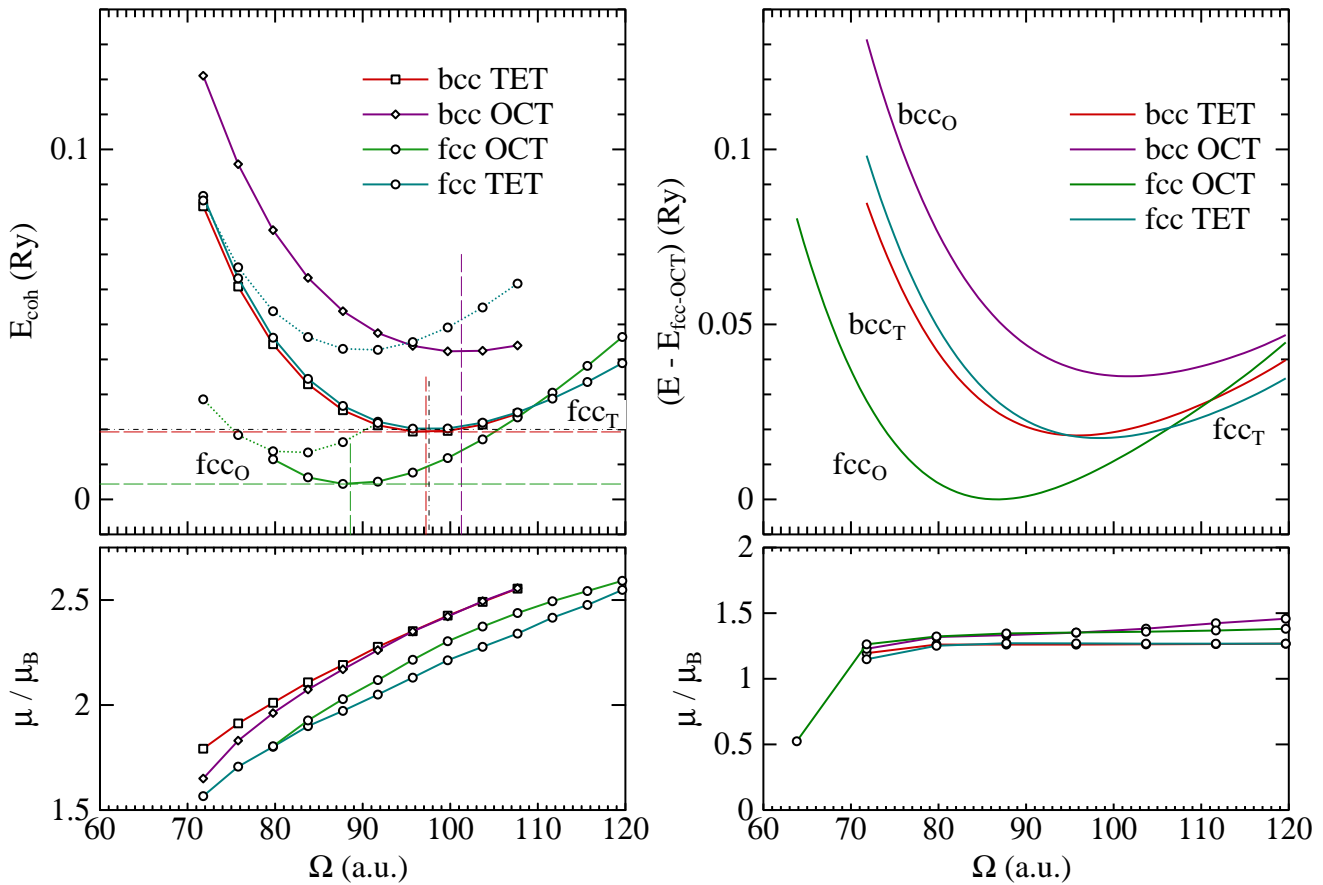}
\end{center}
\end{figure*}

As emphasized before, we will keep the parameters of pure Fe
unchanged as we seek a model for H~in Fe. We will find such a
model by comparison with properties of iron monohydrides of
stoichiometry FeH, that is, the concentrated limit and then test
our model's transferability into the dilute limit. 

In a series of three papers,\cite{CE1,CE2,CE3}
Els\"asser~\ea\ have made a comprehensive study of the compound
FeH in the framework of density functional theory. One is
interested in four putative phases, namely fcc and bcc Fe each
having one H atom in either tetrahedral or octahedral
sites. These are illustrated in Fig.~\ref{fig_Interstitials}; and
Fig.~\ref{fig_FeH-EV} shows energy--volume curves for these four
phases calculated using LSDA-GGA in the full potential LMTO
method\cite{Methfessel00} (see also Fig.~5 in
ref~[\onlinecite{CE1}]).

\begin{figure*}
\caption{\label{fig_FeH-bands} (color online) Energy bands for
  bcc tetrahedral FeH, calculated at the lattice constant of pure
  bcc Fe. The upper panels show majority and the lower minority
  spin states.  The coloring is such that H-$s$ character is red
  and Fe-$d$ character is blue. Fe-$s$ bands are green. The Fermi
  energy is indicated by a horizontal line. Note that the Fe-$4s$
  band has been pushed above the $d$-bands. Bands on the left are
  from our tight binding model and on the right are bands
  calculated in the LSDA-GGA.}
\begin{center}
\includegraphics[scale=1.1]{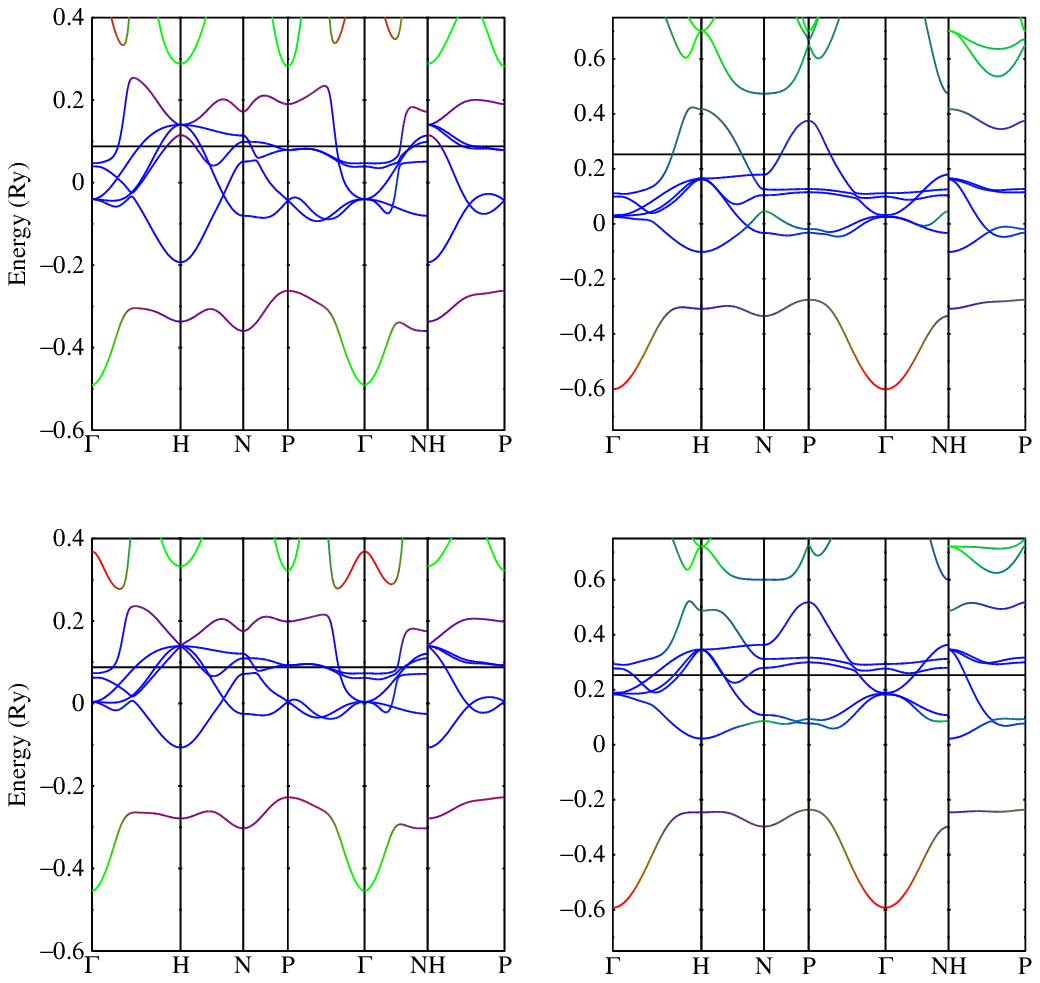}
\end{center}
\end{figure*}

Examination of the upper sketch in Fig.~\ref{fig_Interstitials}
shows that the displacement of the tetrahedral interstitial atom
in the bcc structure towards the octahedral site brings the
impurity atom from above the second neighbor bond, at right
angles until it finally rests at the bond center. This is
precisely the situation envisaged by Haas~\ea\cite{Haas98} in
their proposal of the screening function, and we therefore expect
for a model to be transferable, we will require it to be non
orthogonal. There is also a strong argument for the retention of
the Fe~$4s$ orbital even though, as we have seen, it does not
lead to a significantly better model for pure Fe than the
orthogonal~$d$.\cite{Paxton96} The argument for its inclusion
follows from an examination of Fig.~\ref{fig_FeH-bands} which
shows LSDA-GGA energy bands for bcc tetrahedral FeH. The bands
are colored according to the eigenvector weights coming from
LMTO's from H~$1s$ (red) or Fe~$3d$ (blue). The H~$1s$ band is
split off from the Fe~$3d$ bands and has similar width. The
Fe~$4s$ band which in pure~Fe has its bottom below the Fe~$3d$
bands and which hybridizes with them (see
Fig.~\ref{fig_Fe-bands}) is pushed up above the top of the
Fe~$3d$ bands by repulsion of the H~$1s$ band. This means that
the Fermi energy remains near where it is in pure~Fe. Roughly
speaking one might say that the single $4s$~electron per atom in
pure Fe is transferred to the hydrogen atom to complete its 1$s$
shell, or rather to fill the H~$1s$ band. At first glance it may
seem natural to neglect the Fe~$4s$ bands in FeH. But a
difficulty will arise if we adapt a $d$-only model by adding just
an extra H~$1s$ orbital. Hydrogen brings one electron with it and
to fill the split-off H~$1s$ band an electron will be drawn down
from the Fe~$3d$ bands consequently lowering the Fermi level. If
we were only interested in FeH then we could just adjust $N_d$,
the number of $d$~electrons; but this will introduce an
inconsistency in going to the dilute limit: $N_d$ will somehow
need to be continuously adjusted at Fe~atoms successively further
away from an impurity~H. It is very hard to see how this problem
could be overcome except possibly by alloting {\it two} electrons
to the hydrogen impurity; while it is solved naturally by the
Fe~$4s$ falling back into place as an Fe atom finds itself remote
from the influence of impurity. We emphasize that in the non
orthogonal~$sd$ model and its extension to impurities the number
of electrons is {\it not} a parameter---as long as all occupied
bands are included in the hamiltonian we can happily take the
number of electrons from the periodic table.

Therefore we take over the pure Fe~non orthogonal~$sd$ model and
we add parameters to account for the additional H~$s$ band. We
need Fe--H $\sds$ and $\sss$ hopping and overlap parameters but
we do not require H--H interaction parameters since even the
closest approaching interstitial sites are distant more than
three times the length of the H$_2$ molecular bond. The $\sds$
and $\sss$ integrals establish the {\it width} of the H~$s$ band
while its {\it position} with respect to the $d$~bands is set by
the on-site energy, $\E_s$ of the H~$s$ orbital. We also require
Hubbard-$U$ parameters\cite{Finnis98a,Finnis03} for H and Fe, but
these are not critical and 1.2~Ry and 1~Ry are good
choices. Essentially these lead to approximate charge neutrality
as expected in metals and their alloys.\cite{Pettifor87} For
simplicity we take the Stoner parameter for H~to be
zero. Tetrahedral bcc FeH is ferrimagnetic, both in LSDA and in
our tight binding model, the H~atom carrying a small moment, less
than 1~$\mu_{\text{B}}$ (aligned opposite to that of the Fe~atom
{\it cf.,} Fig.~8 in ref~[\onlinecite{CE1}]).
\begin{table}
\begin{center}
\caption{\label{tbl_parms-FeH} H on-site, and Fe--H interaction
  parameters of our tight binding model. All quantities are given
  in atomic Rydberg units. For all these integrals we use
  $r_1=0.8$ and $r_c=2$ in units of the pure Fe bcc lattice
  constant, $a=2.87$~\AA; for the pair potential we use
  $r_1=0.8$ and $r_c=0.95$ in the same units.}
\begin{tabular}{|c|c|c|}
\hline
  $\E_s-\E_d$   & \multicolumn{2}{c|}{--0.085}  \\
  $ U_{\hbox{Fe}} $   & \multicolumn{2}{c|}{1.0} \\
  $ U_{\hbox{H}}  $   & \multicolumn{2}{c|}{1.2} \\
\hline
  &  &  $q$ \\
\hline
  $h_{ss}$      & --0.35 &  0.776 \\   
  $h_{sd}$      & --0.14 &  0.454 \\   
  $s_{ss}$      & 0.27   &  0.286 \\   
  $s_{sd}$      & 0.22   &  0.473 \\   
\hline
  $B$           & \multicolumn{2}{c|}{299.6}  \\   
  $p$           & \multicolumn{2}{c|}{2.6922} \\   
\hline
\end{tabular}
\end{center}
\end{table}

\begin{figure}
\caption{\label{fig_hopping} (color online) Hopping and
  overlap integrals as functions of bond length, $r$, in the $sd$
  non orthogonal model. Except in the case of $dd\delta$ the
  dotted lines are the overlap integrals corresponding to the
  hopping integrals of the same color. Vertical dotted lines
  indicate the Fe--H bond length in bcc tetrahedral FeH at
  equilibrium volume, and the Fe--Fe bond lengths of the first
  six neighbors in pure bcc Fe.}
\begin{center}
\includegraphics[scale=0.4]{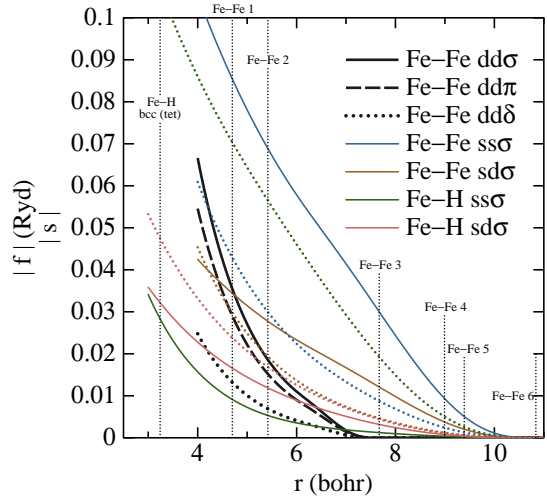}
\end{center}
\end{figure}

To find the additional parameters we have resorted to fitting
these to the four equilibrium atomic volumes and three cohesive
energy differences marked with dashed lines on
Fig.~\ref{fig_FeH-EV}. We do this using Schwefel's multimembered
evolution strategy.\cite{Schwefel77,Schwefel93} For the Fe--H
pair potential we employ
\begin{equation*}
     \phi(r) = \frac{B}{r}\> \e^{-pr}.
\end{equation*} 
The resulting parameters are displayed in
table~\ref{tbl_parms-FeH} and the hopping integrals are shown
graphically in Fig.~\ref{fig_hopping} to illustrate their
relative magnitudes and ranges. In the same figure we show the
hopping integrals for Fe which are, of course, identical to those
of our non orthogonal~$sd$ model of
section~\ref{sec_PureFe}. With reference to our remarks in
section~\ref{subsec_scaling} we note that all our hopping and
overlap integrals have the simple exponential form up to the
distance $r_1$ beyond which they are augmented so as to go
continuously and differentiably to zero at $r_c$. These distances
are not strictly parameters of the model and are not used in the
fitting. They are chosen intuitively; for example one expects
just first neighbors in hcp and fcc, and first plus second
neighbors in the bcc structures to be interacting through $dd$
hopping whereas the $s$~electrons in pure Fe are essentially free
electron like and hence ``do not take kindly to being treated
within a TB framework''.\cite{Pettifor77} They are best
represented by longer ranged interactions. These points are
illustrated in Fig.~\ref{fig_hopping} and the values of $r_1$ and
$r_c$ can be found in tables~\ref{tbl_parms-PureFe}
and~\ref{tbl_parms-FeH}.  The use of fifth degree polynomials to
augment the tails is necessary to acheive a smooth join; it can
lead to small kinks as seen in Fig.~\ref{fig_hopping}, but these
are designed to fall in between neighbor shells and so minimize
their effect. This is why the parameters $r_0$ and $r_c$ are made
to scale with the lattice constant. The resulting energy bands
are plotted in Fig.~\ref{fig_FeH-bands} for comparison with the
LSDA-GGA bands. The resulting energy volume curves are shown in
Fig.~\ref{fig_FeH-EV}. The TB model does not reproduce the
magnetic moments of the LSDA-GGA in Fig.~\ref{fig_FeH-EV}
quantitatively since this is a sensitive function of the density
of states at the Fermi level in the non magnetic crystal and our
energy bands are only in qualitative agreement with the LSDA-GGA.

Table~\ref{tbl_props-FeH} summarizes the equilibrium properties
of the four hydride phases shown in Fig.~\ref{fig_FeH-EV}. The
question of site selectivity, especially in bcc~Fe is important
and we will revisit it in the dilute limit, below, in
section~\ref{subsubsection_edis}.

\begin{center}
\begin{table}
\caption{\label{tbl_props-FeH} Equilibrium volumes per Fe~atom
  and cohesive energies of the four FeH phases following
  evolution optimisation, compared to the target values. Cohesive
  energies are relative to the fcc octahedral (rocksalt)
  phase. The final column shows the radius of the interstitial
  site based on a lattice of hard spheres at the equilibrium
  volume of pure Fe and taken from Leslie.\cite{Leslie} All
  quantities are given in atomic Rydberg units.}
\begin{tabular}{|c|c|c|c|c|c|}
\hline
        & \multicolumn{2}{c|}{TB} & \multicolumn{2}{c|}{Target} & radius\\
\hline
        &    $E_{\hbox{coh}}$ & $\Omega$ &    $E_{\hbox{coh}}$ & $\Omega$ & \\
\hline
 fcc OCT &  0.0  &  86.90  &  0.0    &  88.59 & 0.98 \\
 fcc TET & 0.017 &  98.64  &  0.016  &  97.58 & 0.53 \\
 bcc TET & 0.018 &  96.16  &  0.015  &  97.23 & 0.68 \\
 bcc OCT & 0.035 & 101.75  &  0.038  & 101.28 & 0.36 \\
\hline
\end{tabular}
\end{table}
\end{center}

\section{Predictions of the F\lowercase{e}--H model}
\label{sec_predictions-FeH}

\subsection{Iron hydride}
\label{subsec_FeH}

\begin{figure}
\caption{\label{fig_bcc-int} (color online) Illustrates the
  translations of the bcc intersitials in constructing our
  adiabatic potential surfaces, after the three dimensional
  drawing of Fig.~1 in Krimmel~\ea\cite{Krimmel94b} The figure
  represents an (010) face of the bcc lattice with Fe atoms as
  black circles at each corner. The octahedral sites are shown as
  squares, the central, filled one being the one occupied in
  octahedral FeH. Of the four tetrahedral sites, shown as
  triangles, one is occupied in tetrahedral FeH and this is shown
  filled in here. The point, S, is midway between two tetrahedral
  sites---the expected diffusion path of H in Fe\cite{Hirth80}
  which is highlighted in red here and in
  Fig.~\ref{fig_AdPot-bcc}. Those displacements which are in the
  (010) plane are indicated.}
\begin{center}
\includegraphics[scale=0.5]{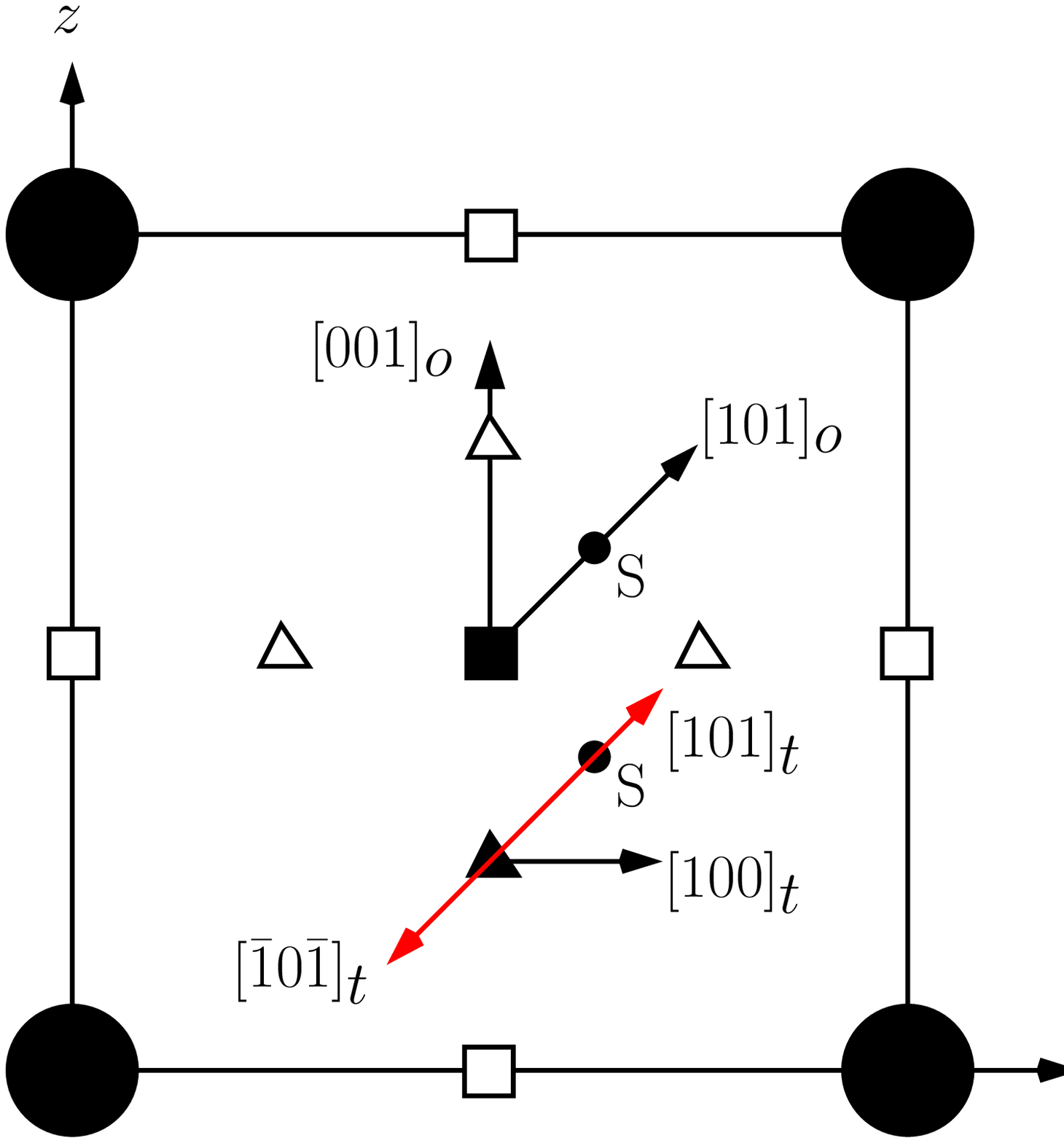}
\end{center}
\end{figure}

\begin{figure*}
\caption{\label{fig_AdPot-bcc} (color online) Adiabatic potential
  surface sections of bcc FeH: left LDA,\cite{CE3} right
  TB. These curves show the energy as a function of the
  displacement of the H~sublattice relative to the
  Fe~sublattice. The curves which start at the point ``O'' refer
  to displacements from the octahedral site phase; a H~atom
  initially at position $[\sh 0\sh]$ translates in the directions
  indicated. Along the $[001]$ direction it eventually falls into
  a vacant tetrahedral site (see Fig.~\ref{fig_bcc-int}). This
  curve hence represents the transition to the tetrahedral-site
  phase. Translation along $[101]$ takes the H~atom to a position
  midway between two, vacant, tetrahedral sites---this point is
  marked ``S''. For a H~atom initially occupying a tetrahedral
  site, translation along $[101]$ moves it to an adjacent,
  unoccupied, tetrahedral site, the half-way point being the same
  point ``S''. The translation labels are vectors referred to
  Fig.~\ref{fig_bcc-int}. For each case, LDA and TB, the
  calculations are at fixed atomic volume, namely the equilibrium
  volume of the bcc tetrahedral phase of FeH, see
  table~\ref{tbl_props-FeH}; $a_0$ is the corresponding
  equilibrium lattice constant.}
\begin{center}
\includegraphics[scale=1]{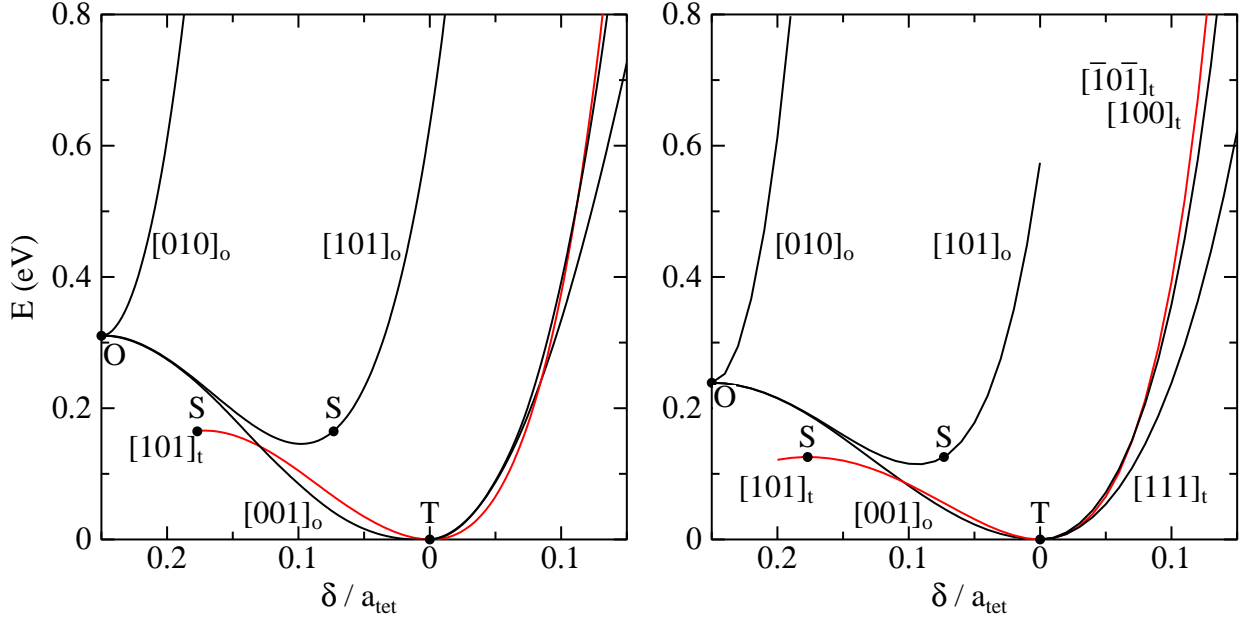}
\end{center}
\end{figure*}

\begin{figure*}
\caption{\label{fig_AdPot-fcc} Adiabatic potential surface
  sections of fcc FeH: left LDA,\cite{CE3} right TB. At the point
  ``O'' we have the rocksalt phase, from which translation of the
  H~sublattice along a $\langle 111\rangle$ direction transforms
  the structure to the zincblende phase in which tetrahedral
  sites are occupied. The energy maximum between ``O'' and ``T''
  is located close to where the H~atom squeezes between an
  equilateral triangle of Fe~atoms in the (111) plane. At the
  maximum along $\langle 110\rangle_o$, and along $\langle
  001\rangle_t$, the H is positioned mid-way between two nearest
  neighbor Fe atoms (see Fig.~1 of
  ref~[\onlinecite{Krimmel94a}]). Note that both these two energy
  barriers are predicted by the TB model with quantitative
  accuracy. The calculations are at the calculated equilibrium
  volume of the fcc octahedral (rocksalt) phase of FeH, see
  table~\ref{tbl_props-FeH}; $a_0$ is the corresponding
  equilibrium lattice constant.}
\begin{center}
\includegraphics[scale=1]{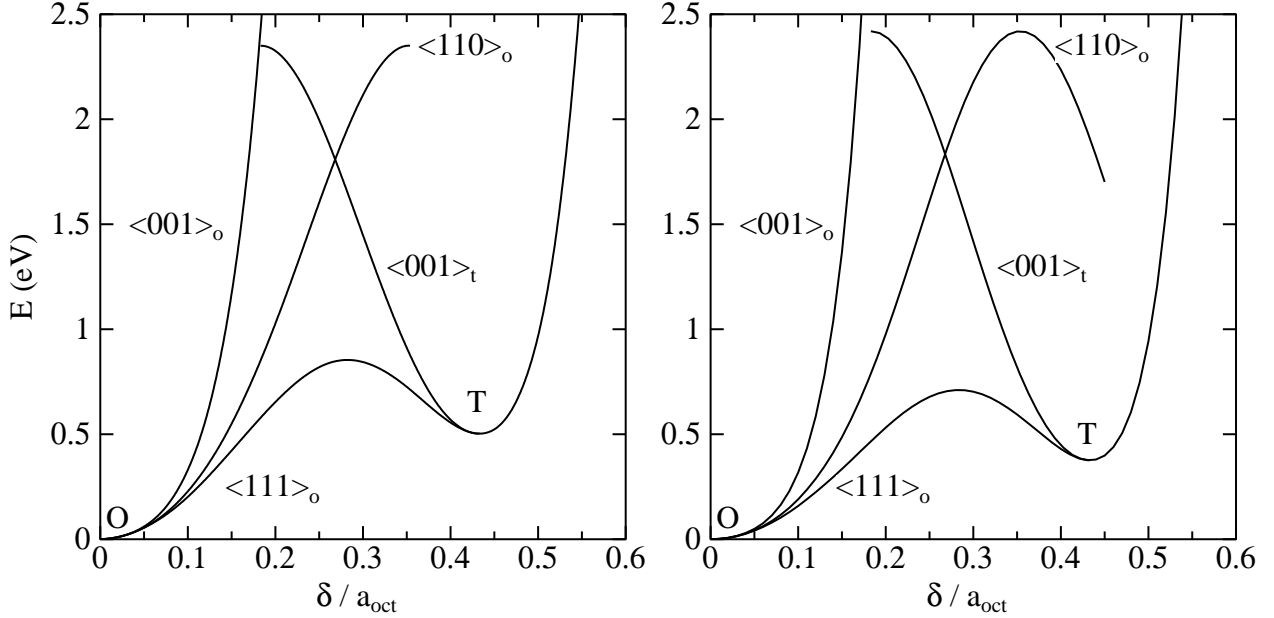}
\end{center}
\end{figure*}

Our first test of the tight binding model is to compare the
resulting adiabatic potential surface section with the results of
calculations by Els\"asser~\ea\cite{CE3,Krimmel94b} which were
made in the local density approximation (LDA) to DFT. In these
calculations the H~sublattice is displaced with respect to the
Fe~sublattice in both bcc and fcc FeH in a chosen set of
directions so as to explore the curvatures and barriers of the
potential energy landscape. For the case of the bcc structure,
Fig.~\ref{fig_bcc-int} shows some of the displacment paths. The
potential sections from previous LDA\cite{CE3} and our present
tight binding model are shown in
Fig.~\ref{fig_AdPot-bcc}. Whereas the relative energies of the
tetrahedral and octahedral sites have been established by the
fitting, the remainder of the these curves amount to {\it
  predictions} of the tight binding model. They turn out to be be
in remarkable, quantitative agreement with the LDA calculations
in the bcc and fcc case, the latter being shown in
Fig.~\ref{fig_AdPot-fcc}. These curves exploit to the full the
notion discussed in section~\ref{subsec_trans}, above, of
environment dependent screening of hopping integrals as the
hydrogen approaches Fe--Fe first and second neighbor bonds and
indeed penetrates the bond to lie directly in between the two
atoms. It is exactly in this situation that one expects the
Fe--Fe bond integrals to be strongly modified by screening, and
clearly our model captures this well in a non orthogonal two
center description. In particular note, in reference to
Fig.~\ref{fig_AdPot-bcc} that the minimum energy (saddle) point
along the $[101]_o$ path lies to the {\it left} of the
point~``S'' in both LDA and in our TB model. This implies that
the $[101]_t$ minimum energy diffusion path in reality is {\it
  bowed} slightly towards the center of
Fig.~\ref{fig_bcc-int}. The strongest test of the environment
dependence however is in the fcc hydrides of
Fig.~\ref{fig_AdPot-fcc}. The energy barrier at the maximum of
the $\langle 110\rangle_o$ path, coinciding with the maximum of
the $\langle 001\rangle_t$ path is perfectly rendered by the TB
model without having been fitted and this corresponds to the
extreme instance of screening in which the H~atom becomes
positioned at the center of the first neighbor Fe--Fe bond (see
Fig.~1, ref~[\onlinecite{Krimmel94a}]).

\subsection{H in F\lowercase{e}---the dilute limit}
\label{subsec_FeH-dilute}

We concentrate on three predicted properties of iron in this
section. First is the {\it dissolution energy}\cite{Jiang04} or
zero temperature heat of solution of hydrogen in Fe. Included in
this study is the matter of the site selectivity. Second is the
{\it binding energy}\cite{EAC09} or 0$^{\circ}$K segregation
energy of H~to the (001) surface of Fe. Third, and of great
importance to the question of hydrogen embrittlement, is the
binding of H~atoms to a vacancy in Fe. 

\subsubsection{Dissolution energy}
\label{subsubsection_edis}

Following Ramasubramaniam~\ea\cite{EAC09} we construct a 54~atom
supercell as we did in section~\ref{subsubsec_PureFe-vacancies} and
whose total energy we denoted $E(\text{Fe}_{54})$. We then place a
hydrogen atom at either a tetrahedral or an octahedral site and
minimize the total energy by relaxation. The resulting total
energies are denoted $E(\text{Fe}_{54}H)$. We do not allow the
volume to relax. Then the dissolution energy is\cite{Jiang04}
\begin{equation} 
\label{eq_edis}
\Edis=E(\text{Fe}_{54}\text{H})-E(\text{Fe}_{54})-\half E_{{\text{H}}_2}
\end{equation} 
Our model does not contain H--H interactions, but {\it faux de
  mieux} we may take $E_{{\text{H}}_2}=-4.75$~eV from experiment
or from quantum chemistry.\cite{Skinner91,Jiang04}
For each of the three calculations we employ a mesh of $12\times
12\times 12$ ${\bf k}$-points and use first order generalized
Gaussian integration of the Brillouin zone with a width of
2.5~mRy.\cite{Methfessel89} Results are shown in
table~\ref{tbl_Edis}. These are in remarkably good quantitative
agreement with both observations and LSDA-GGA calculations. In
particular we predict the tetrahedral site to be preferred over
the octahedral, as is well established.\cite{Hirth80} We may
point out here that this is not a trivial result: carbon in
contrast, while preferring the tetrahedral site in the ficticious
bcc-based carbide, transfers to the octahedral site in the dilute
limit.\cite{Paxton10} In the effective medium theory, upon which
the embedded atom potentials (EAM) are based, H~prefers the
octahedral site.\cite{Puska84}

\begin{center}
\begin{table}
\caption{\label{tbl_Edis} Dissolution energy, in eV, of H~in
  Fe in both tetrahedral (TET) and octahedral (OCT)
  interstices. Present results are marked TB, experimental and
  LSDA-GGA values are taken from Jiang and Carter.\cite{Jiang04}}
\begin{tabular}{|c|c|c|}
\hline
     &   TET     &  OCT   \\
\hline
 TB  & 0.273   &  0.354  \\
expt.& 0.296   &         \\
 GGA & 0.19   &  0.32  \\
\hline
\end{tabular}
\end{table}
\end{center}

\subsubsection{H segregation to the (001) surface of Fe}
\label{subsubsection_bind}

\begin{figure}
\caption{\label{fig_001Surface} (color online) Three possible
  binding sites of H on the (001) surface of Fe, after
  Ramasubramaniam~\ea\cite{EAC09} Four large circles represent
  the Fe atoms at the corners of a unit cell of the (001) face of
  the bcc lattice. At the center is the ``hollow'' site, a
  smaller circle; this may be displaced along [100] by an amount
  $\delta$ to become the ``quasi-threefold'' (QT) site indicated
  by a triangle. The ``bridge'' site is shown as a square.  It is
  important to recognize that the bridge and hollow sites in the
  plane of the truncated bulk surface are octahedral interstices,
  whereas the QT site at $\delta=0.25a_0$ is a tetrahedral
  site. If a H~atom at the bridge site is displaced up or down by
  $0.25a_0$ then it comes to occupy a tetrahedral site. Here,
  $a_0=2.87$~\AA\ is the equilibrium pure $\alpha$-Fe lattice
  constant.}
\begin{center}
\includegraphics[scale=0.5]{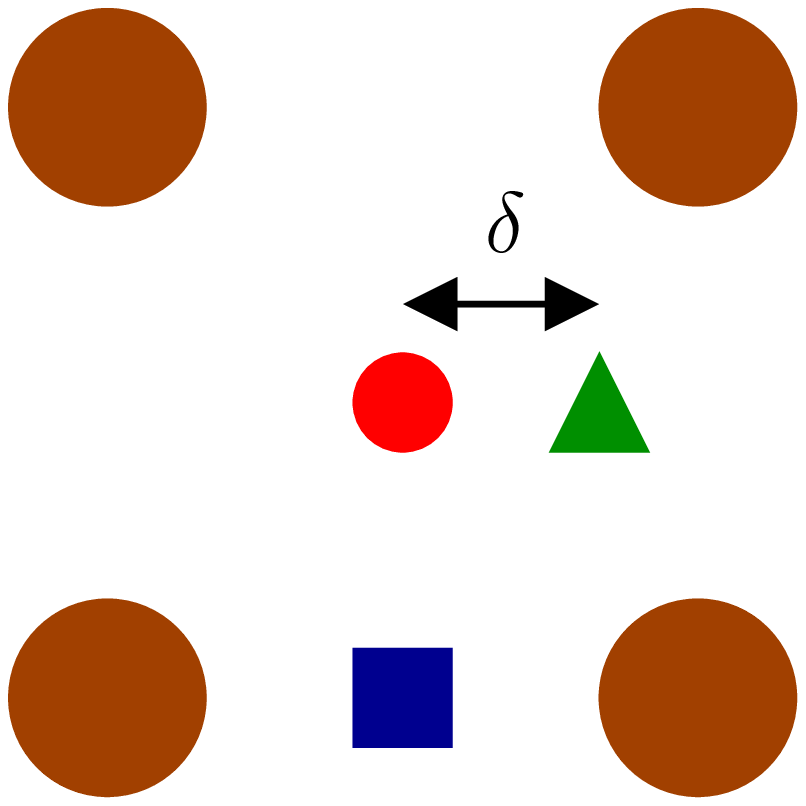}
\end{center}
\end{figure}

Three binding sites of H~to the (001) surface of Fe have been
identified.\cite{EAC09} These are illustrated in
Fig.~\ref{fig_001Surface}. We have constructed
supercells of $2\times 2\times 5$ cubic two-atom unit cells with
three layers of vacuum inserted along the long axis. The slab
contains 40~Fe atoms and the total energy of the fully relaxed
supercell is denoted $E^{\text{surf}}(\text{Fe}_{40})$. We place
one H~atom at one of the three adsorption sites in
Fig.~\ref{fig_001Surface} and relax the structure by energy
minimization. Allowing all atoms to relax we denote the total
energy $E^{\text{surf}}(\text{Fe}_{40}H)$.  The associated
``adsorption energy'' is\cite{EAC09}
\begin{equation*} 
\Eads=E(\text{Fe}_{54}\text{H})-E(\text{Fe}_{54})-\half E_{{\text{H}}_2}
\end{equation*} 
and by combining the previous two equations the ``binding energy'' is\cite{EAC09}
\begin{equation} 
\label{eq_ebind}
\Ebind = \Edis^t - \Eads
\end{equation} 
in which the reference energy, or chemical potential, of gaseous
H$_2$ has canceled. $\Edis^t$ is the dissolution
energy~(\ref{eq_edis}) at a tetrahedral site
(table~\ref{tbl_Edis}). We have calculated the three quantities
using a $12\times 12\times 1$ ${\bf k}$-point mesh and the same
Brillouin zone integration as above. In
table~\ref{tbl_001Surface} we show our calculated binding
energies, the displacement $\delta$ in Fig.~\ref{fig_001Surface}
and the height, $h$, from the (001) surface constructed as the
difference in $z$-coordinates of the H~atom and the average from
the four topmost Fe~atoms.

\begin{center}
\begin{table}
\caption{\label{tbl_001Surface} Predicted structure and
  energetics of H~adsorbed on Fe~(001). We show for the QT,
  hollow (H) and bridge (B) sites of Fig.~\ref{fig_001Surface} the
  displacent $\delta$ and height, $h$, above the surface (all
  in~\AA) and the 0$^{\circ}$K segregation or binding energy,
  $\Ebind$, in eV. In parentheses are the LSDA-GGA results of
  Ramasubramaniam~\ea\cite{EAC09}}
\begin{tabular}{|c|cc|cc|cc|}
\hline
 & \multicolumn{2}{c|}{$\delta$} & \multicolumn{2}{c|}{$h$} & \multicolumn{2}{c|}{$\Ebind$} \\
\hline
 &  TB & (GGA) & TB & (GGA) & TB & (GGA) \\
\hline
 QT  & 0.635 & (0.19) & 0.31 & (0.38) & 0.241 & (0.768) \\
 H   &       &        & 0.27 & (0.38) & 0.191 & (0.775) \\
 B   &       &        & 0.85 & (1.20) & 0.222 & (0.655) \\
\hline
\end{tabular}
\end{table}
\end{center}

The predictions of our model are only in reasonable agreement
with the LSDA-GGA.\cite{EAC09} The heights above the surface are
well rendered; the displacement, $\delta$, is significantly
larger, but is consistent with the preference for tetrahedral
site occupancy. As we point out in the caption to
Fig.~\ref{fig_001Surface}, $\delta=0.25a_0=0.71$~\AA\ puts the
H~atom into a surface tetrahedral site and our model does exactly
that; in contrast the LSDA-GGA quite surprisingly results in a
much smaller $\delta$. In the same vein, the height of the H~atom
above the bridge site, 0.85~\AA, is close to $0.25a_0$, and we
find another local minimum at 0.34~\AA\ below the bridge
site. Thus the strongest binding in the TB model is to surface
tetrahedral sites and the surface octahedral site is indeed not a
local energy minimum. In this way the binding energies are in
poor agreement with the LSDA-GGA and may reflect the limitations
in transferability (section~\ref{subsec_trans}) in that the model
retains its bulk-like features at the surface. $\Ebind$ is in
fact the 0$^{\circ}$K segregation energy, usually defined as the
energy needed to remove the impurity from the surface and place
it into the interior of the crystal. The LSDA-GGA shows the
smallest adsorption energy (largest $\Ebind$) to be at the hollow
site; whereas we find it at the QT site and at this coverage this
is {\it not} consistent with experiment which shows a transition
at 100$^{\circ}$K from hollow to QT site selectivity between
about 0.3ML and 1ML,\cite{Merrill96} while our calculations and
the LSDA-GGA\cite{EAC09} are at 0.25ML.

Both the QT and bridge sites are at local minima in the potential
energy in our model. This is consistent with the
LSDA-GGA.\cite{EAC09} However the hollow site is a local saddle
point having an almost flat energy surface with respect to
small displacements parallel to the surface; if we
displace the H~atom a sufficient amount then the structure
relaxes into the QT site occupancy. This is inconsistent with the
LSDA-GGA in which surprisingly, in view of there being another
local minimum at QT just 0.19\AA\ distant, the hollow site is at a
local minimum.\cite{EAC09}

To some extent our choice of chemical potential for $H_2$,
$E_{{\text{H}}_2}$, is arbitrary; however the observed bond
energy leads to a very good rendering of the 0$^{\circ}$K heat of
solution (dissolution energy) of H~in Fe,
table~\ref{tbl_Edis}. On the other hand it leads to a positive,
but small, adsorption energy, $\Eads$, which means that in our
model $H_2$ will not dissociate on the (001) surface of~Fe. In
order to model the surface adsorption properly we could make an
{\it ad hoc} adjustment of $E_{{\text{H}}_2}$. This would be at
the expense of less accurate $\Edis$. For example, if we used the
Skinner and Pettifor tight binding model of
hydrogen,\cite{Skinner91} then we'd have
$E_{{\text{H}}_2}=-4.30$~eV rather than $-4.75$~eV. In that case
our dissolution energy in the tetrahedral site becomes 0.05~eV
(rather than 0.27~eV, {\it cf\/} table~\ref{tbl_Edis}) but the
adsorption energies are then negative as they should be. Of
course the segregation energies (table~\ref{tbl_001Surface})
remain unchanged by this redefinition of the hydrogen chemical
potential.

\subsubsection{H segregation to a vacancy in Fe}
\label{subsubsection_vacancy}

It is believed that the trapping of H~to vacancies in Fe is of
central importance in the effects of H~on mechanical
behavior.\cite{Hirth80,Takai08,Kirchheim10} It is also known that
dissolved hydrogen results in a dramatic increase in the vacancy
concentration in several metals including
Fe,\cite{Iwamoto99,Fukai03} caused through segregation induced
lowering of the vacancy formation enthalpy.\cite{Kirchheim07b} We
can show that our model is able to demonstrate these facts by
comparison with LSDA-GGA calculations of the 0$^{\circ}$K
segregation energy, $\Ebindv$, of up to seven H~atoms to a single
vacancy in Fe.\cite{EAC09,Tateyama03} The principal result, which
we also predict in our TB model is that up to five H~atoms may
bind to a vacancy with a positive segregation energy, but the
sixth has a small negative $\Ebindv$. Here we follow Tateyama and
Ohno\cite{Tateyama03} and Ramasubramaniam~\ea\cite{EAC09} and
define $\Ebindv$ as the 0$^{\circ}$K segregation energy of a
H~atom from a bulk tetrahedral site to a vacancy to which $(n-1)$
H~atoms are {\it already segregated.} Hence we set up a 53~atom
supercell as in section~\ref{subsubsec_PureFe-vacancies}; then in
reference to figure~5 in Ramasubramaniam~\ea,\cite{EAC09} if the
vacant site is at $[\half\half\half]$ in the bcc supercell we add
H~atoms successively in (1)~$[\half\half 0]$, (2)~$[\half\half
  1]$, (3)~$[\half 1\half]$, (4)~$[\half 0\half]$,
(5)~$[1\half\half]$, and (6)~$[0\half\half]$ octahedral
interstices---these are the centers of the six \{001\} faces
bounding the vacant site.  Finally a seventh H~atom may be placed
at the vacant site. These supercells are relaxed by energy
minimization and we denote the total energy of the supercell by
$E({\text{Fe}}_{53}{\text{H}}_n)$. Then we
have\cite{Tateyama03,EAC09,Footnote} in analogy with~(\ref{eq_ebind})
\begin{equation*} 
\Ebindv=\Edis^t-\left(E({\text{Fe}}_{53}{\text{H}}_n)-E({\text{Fe}}_{53}{\text{H}}_{n-1})-\half E_{{\text{H}}_2}\right)
\end{equation*} 
which is independent of the chemical potential of~H.
Table~\ref{tbl_FeHvac}\nocite{Tateyama03} shows our segregation
energies, compared to LSDA-GGA. The relaxation pattern is very
simple in all cases except $n=3$ and $n=5$. In the simple
instances, each H~atom relaxes perpendicularly to its \{001\}
face, by an amount we denote $\delta^{\text{even}}_{\perp}(n)$,
towards the vacant site. The displacement decreases as $n$
increases both in LSDA-GGA\cite{Tateyama03} and our TB model. In
each of the cases $n=3$ and $n=5$ there is {\it one} H~atom which
follows this trend whereas the remaining $(n-1)$ H~atoms are
displaced both towards the vacancy by
$\delta^{\text{odd}}_{\perp}(n)$ {\it and}, by an amount
$\delta_{\parallel}(n)$ in a direction parallel to the \{001\}
face containing the site where the H~atom was originally placed,
in a $\langle 100\rangle$ direction.

\begin{center}
\begin{table}
\caption{\label{tbl_FeHvac} Segregation of H~atoms to a vacancy
  in Fe. We show our model's predicted $\Ebindv$ compared to
  LSDA--GGA results,\cite{Tateyama03} quoted by
  Ramasubramaniam~\ea,\cite{EAC09} in eV. Also shown are the
  displacements of the H~atom towards the vacancy, and away from
  the octahedral site in the \{001\} plane in which it was
  originally placed. In cases of higher symmetry the displacement
  of all H~atoms is an amount $\delta^{\text{even}}_{\perp}(n)$
  normal to the \{001\} face and towards the vacant site. In the
  cases $n=3$ and $n=5$ {\it one} atom follows this displacement,
  while all those remaining move both perpendicular to the
  face---by an amount $\delta^{\text{odd}}_{\perp}(n)$---and
  parallel to the face in a $\langle 100\rangle$ direction by an
  amount $\delta_{\parallel}(n)$, rather like the knight's move
  in chess. A displacement
  $\delta_{\parallel}=0.25a_0=0.71$~\AA\ will take the H~atom
  into a tetrahedral site. Displacements are given in \AA.}
\begin{tabular}{|c|c|c|c|c|c|c|}
\hline
  $n$   &  \multicolumn{2}{c|}{$\Ebindv$} &
$\delta^{\text{even}}_{\perp}(n)$ & expt.\footnotemark[1] & $\delta^{\text{odd}}_{\perp}(n)$ & $\delta_{\parallel}(n)$ \\
\hline
        &    TB  & LSDA-GGA  & & & &\\
\hline
  1     &  0.319  &  0.559  &  0.25  &$0.4\pm 0.1$&& \\
  2     &  0.330  &  0.612  &  0.27  &&& \\
  3     &  0.263  &  0.399  &  0.19  && 0.27 & 0.35  \\
  4     &  0.160  &  0.276  &  0.28  &&& \\
  5     &  0.144  &  0.335  &  0.13  && 0.26 & 0.25  \\ 
  6     &--0.033  &--0.019  &  0.19  &&& \\
  7     &--0.474  &--2.68   &  0.14  &&& \\
\hline
\end{tabular}
\footnotetext[1]{Reference [\onlinecite{Myers79}]}
\end{table}
\end{center}

Table~\ref{tbl_FeHvac} shows very much better agreement with the
LSDA-GGA than in the case of surface segregration. This probably
reflects the better transferability into the less
undercoordinated environment. Our absolute values of $\Ebindv$
are no more than 50\% underestimated while the trends are in
perfect accord: we observe the increase in segegration energy
going from $n=1$ to $n=2$ implying that a H~atom segregates more
readily to a vacancy {\it that has already trapped} a H~atom. We
also see that up to five H~atoms will segregate exothermally to a
vacancy, while the sixth segregates endothermically. The
displacement patterns in the symmetric cases are consistent in
magnitude with the LSDA-GGA\cite{Tateyama03} and follow the trend
of decreasing $\delta^{\text{even}}_{\perp}$ with increasing
$n$. For the case $n=1$ we obtain
$\delta^{\text{even}}_{\perp}=0.25$~\AA\ which agrees well with
the LSDA-GGA calculated value of 0.22~\AA.\cite{Tateyama03} An
experimental estimate of $0.4\pm 0.1$~\AA\ was obtained for
deuterium in Fe by ion channeling.\cite{Myers79} Effective medium
theory for $n=1$ results in $\delta^{\text{even}}_{\perp}=0.5\pm
0.1$~\AA\ in~Fe\cite{Norskov82} and $0.46\pm
0.07$~\AA\ in~Nb.\cite{Cizek04} The octahedral sites in which the
H~atoms are originally placed correspond to the hollow sites at
the (001) surface, and as in the surface case the atoms relax
into the vacuum or vacancy and, symmetry permitting, laterally
towards the tetrahedral positions. The interpretation of Tateyama
and Ohno\cite{Tateyama03} that there is an electrostatic
repulsion between H~atoms is unconvincing to us, since we imagine
that this will be screened by the electrons in the vacant
site. We note that in the highly endothermic segregation of a
seventh H~atom to the vacancy there is still an inward
relaxation, at least in our model, towards the vacant site, now
occupied by a H~atom. However our $E^v_{\text{bind}}(7)$  is more
than five times smaller than in LSDA-GGA.\cite{Tateyama03}

We should note, as Kirchheim has pointed out,\cite{Kirchheim10}
that the reduction in enthalpy of the impurity by segregating to
a defect is entirely equivalent to a reduction in the defect's
enthalpy of formation. Hence ours and the LSDA-GGA binding
energies of table~\ref{tbl_FeHvac} are consistent with the
observed ``superabundant vacancy formation'' in many metals
subject to a high hydrogen fugacity\cite{Iwamoto99,Fukai03} (see
Fig.~7, ref~[\onlinecite{Tateyama03}]).

\section{Discussion and conclusions}
\label{sec_discussion}

We have described simple and robust tight binding models for pure
Fe, transferable from bcc into hcp and fcc structures and hence
able to describe the common phases of Fe, $\alpha$, $\gamma$ and
$\epsilon$. Furthermore we have included a description of the
electronic structure of monohydrides and this model has been
shown to be transferable into the dilute limit of interstitial
H~impurity in Fe. A simple orthogonal $d$-band model is expected
to be most appropriate for the pure transition metals and their
alloys\cite{Pettifor79,Williams80,Pettifor87} and indeed the
addition of $s$~or $p$~electrons does not usually result in
better energetics.\cite{Paxton96,Paxton08} This is confirmed here
(table~\ref{tbl_props-PureFe}) in the case of bulk elastic
constants and structural energy difference. The only improvement
to bulk properties arising from the non orthogonal~$sd$ model is
an improved cohesive energy. Vacancy formation and surface
energies are somewhat improved in the non orthogonal $sd$~model.

The focus on transferability is made in section~\ref{sec_FeH}
where, while not permitting the parameters of the pure Fe model
to be adjusted, we seek additional parameters to describe Fe--H
interactions. We give reasons in section~\ref{sec_FeH} in
addition to the transferability arguments for choosing to extend
the non orthogonal~$sd$ rather than the orthogonal~$d$ model to
the description of hydrogen. There are only few additional
parameters needed (table~\ref{tbl_parms-FeH}) and we emphasize
that these were fitted to just seven fiducial points in the
LSDA-GGA energy--volume curves for four putative iron hydrides
(Fig.~\ref{fig_FeH-EV}). Possibly as a consequence of our
adoption of a non orthogonal model both for pure Fe and Fe--H,
our resulting model predicts calculated adiabatic potential
surfaces with quantitative accuracy. It is particularly notable
that in these tests H~atoms are brought perpendicularly towards
Fe--Fe bonds to the point that the H~atom comes between the two
host atoms. This happens in both bcc and fcc hydrides; in the
latter case a H~atom also pushes through the triangle of nearest
neighbour Fe atoms in the (111) plane and the matching to the LDA
is excellent (figs.~\ref{fig_AdPot-bcc} and~\ref{fig_AdPot-fcc}).

Our approach has been to find a model purely by reference to the
concentrated limit of a stoichiometric monohydride, FeH, and then
to {\it test} that model into the dilute limit of H~in
Fe. Therefore all the results in section~\ref{subsec_FeH-dilute}
are predictions. In contrast, in constructing a classical model
Ramasubramaniam~\ea\cite{EAC09} needed to put all the properties
that we describe in section~\ref{subsec_FeH-dilute} into the
training set for the potential. In consequence, the tight binding
approach cannot hope to reproduce the quantitative accuracy that
is achieved by a well fitted classical model. However dissolution
energies, site selectivity and vacancy segregation are very well
rendered in the model. Its most obvious shortcoming is in the
prediction of adsorption energies of H~on the (001) surface
of~Fe. The absolute cohesive energy is problematic in
LSDA,\cite{Heggie93,Philipsen96} but even more so in tight
binding (see the caption to table~\ref{tbl_props-PureFe}).
Possibly for this reason we find that H$_2$ will not dissociate
on the (001) surface if we use the known binding energy of the
H$_2$ molecule as our reference. In future work we will need to
account for molecular hydrogen and this matter will be
revisited. On the other hand qualitatively the TB model gives a
reasonable account of H~adsorption which is certainly a subtle
and complex problem in surface physics. In this way the TB
model does not transfer faultlessly into the problem of surface
energetics. Our predictions of segregation to a vacancy, in
contrast, are in very good accord with the known theoretical
LSDA-GGA results and experimental facts. In particular we predict
that a vacancy will bind up to five H~atoms exothermically and
that the segregation energy is somewhat larger to a vacancy at
which one H~atom is already bound. The trapping of vacancies is
central to the mechanism of the action of H~on the mechanical
properties of Fe~alloys.\cite{Hirth80,Takai08,Kirchheim10}

In conclusion, the quantum mechanical tight binding approximation
lies between the first principles LSDA and the atomistic
classical approach to defect energetics in iron. Because the TB
approximation is grounded in electronic structure theory it may
be applied to this question rather easily and just a few
parameters---adjustable within intuitive limits---are
required. Because of this and because of its simplicity the TB
approach may give rise to a better understanding than the LSDA,
which after much labor produces a total energy and force, often
without clear insight to their origins. In contrast the huge
number of parameters and the rather opaque functional form of the
interatomic interactions in the classical potentials, while able
to model many properties quantitatively, must be at risk of
failure once they are transferred into situations for which they
were not fitted. Therefore we expect the TB approximation to
provide a useful and complementary tool to the classical
potentials, and once augmented with parameters to describe
carbon, to become competitive in the atomistic simulation of the
properties of iron and steel.

\section*{Acknowledgments}
We thank Professor P.~Gumbsch for enlightening discussions and
comments on the manuscript.

We are grateful to the Royal Society for the award of
an International Joint Project, JP0872832.

A.~T.~P. is grateful to the German Research Foundation (DFG
project Gu~367/30).

Financial support from the German Federal Ministry for Education
and Research (BMBF) to the Fraunhofer IWM for C.~E. (grant number
02NUK009C) is gratefully acknowledged.


\begin{thebibliography}{90}%
\makeatletter
\providecommand \@ifxundefined [1]{%
 \@ifx{#1\undefined}
}%
\providecommand \@ifnum [1]{%
 \ifnum #1\expandafter \@firstoftwo
 \else \expandafter \@secondoftwo
 \fi
}%
\providecommand \@ifx [1]{%
 \ifx #1\expandafter \@firstoftwo
 \else \expandafter \@secondoftwo
 \fi
}%
\providecommand \natexlab [1]{#1}%
\providecommand \enquote  [1]{``#1''}%
\providecommand \bibnamefont  [1]{#1}%
\providecommand \bibfnamefont [1]{#1}%
\providecommand \citenamefont [1]{#1}%
\providecommand \href@noop [0]{\@secondoftwo}%
\providecommand \href [0]{\begingroup \@sanitize@url \@href}%
\providecommand \@href[1]{\@@startlink{#1}\@@href}%
\providecommand \@@href[1]{\endgroup#1\@@endlink}%
\providecommand \@sanitize@url [0]{\catcode `\\12\catcode `\$12\catcode
  `\&12\catcode `\#12\catcode `\^12\catcode `\_12\catcode `\%12\relax}%
\providecommand \@@startlink[1]{}%
\providecommand \@@endlink[0]{}%
\providecommand \url  [0]{\begingroup\@sanitize@url \@url }%
\providecommand \@url [1]{\endgroup\@href {#1}{\urlprefix }}%
\providecommand \urlprefix  [0]{URL }%
\providecommand \Eprint [0]{\href }%
\@ifxundefined \urlstyle {%
  \providecommand \doi  [0]{\begingroup \@sanitize@url \@doi}%
  \providecommand \@doi [1]{\endgroup \@@startlink {\doibase
  #1}doi:\discretionary {}{}{}#1\@@endlink }%
}{%
  \providecommand \doi  [0]{doi:\discretionary{}{}{}\begingroup
  \urlstyle{rm}\Url }%
}%
\providecommand \doibase [0]{http://dx.doi.org/}%
\providecommand \Doi [0]{\begingroup \@sanitize@url \@Doi }%
\providecommand \@Doi  [1]{\endgroup\@@startlink{\doibase#1}\@@Doi}%
\providecommand \@@Doi [1]{#1\@@endlink}%
\providecommand \selectlanguage [0]{\@gobble}%
\providecommand \bibinfo  [0]{\@secondoftwo}%
\providecommand \bibfield  [0]{\@secondoftwo}%
\providecommand \translation [1]{[#1]}%
\providecommand \BibitemOpen [0]{}%
\providecommand \bibitemStop [0]{}%
\providecommand \bibitemNoStop [0]{.\EOS\space}%
\providecommand \EOS [0]{\spacefactor3000\relax}%
\providecommand \BibitemShut  [1]{\csname bibitem#1\endcsname}%
\bibitem [{\citenamefont {Ramasubramaniam}\ \emph {et~al.}(2009)\citenamefont
  {Ramasubramaniam}, \citenamefont {Itakura},\ and\ \citenamefont
  {Carter}}]{EAC09}%
  \BibitemOpen
  \bibfield  {author} {\bibinfo {author} {\bibfnamefont {A.}~\bibnamefont
  {Ramasubramaniam}}, \bibinfo {author} {\bibfnamefont {M.}~\bibnamefont
  {Itakura}}, \ and\ \bibinfo {author} {\bibfnamefont {E.~A.}\ \bibnamefont
  {Carter}},\ }\href@noop {} {\bibfield  {journal} {\bibinfo  {journal} {Phys.
  Rev. B},\ }\textbf {\bibinfo {volume} {79}},\ \bibinfo {eid} {174101}
  (\bibinfo {year} {2009})}\BibitemShut {NoStop}%
\bibitem [{\citenamefont {Mrovec}\ \emph {et~al.}(2009)\citenamefont {Mrovec},
  \citenamefont {Els{\"a}sser},\ and\ \citenamefont {Gumbsch}}]{Mrovec09}%
  \BibitemOpen
  \bibfield  {author} {\bibinfo {author} {\bibfnamefont {M.}~\bibnamefont
  {Mrovec}}, \bibinfo {author} {\bibfnamefont {C.}~\bibnamefont
  {Els{\"a}sser}}, \ and\ \bibinfo {author} {\bibfnamefont {P.}~\bibnamefont
  {Gumbsch}},\ }\href@noop {} {\bibfield  {journal} {\bibinfo  {journal} {Phil.
  Mag.},\ }\textbf {\bibinfo {volume} {89}},\ \bibinfo {pages} {3179} (\bibinfo
  {year} {2009})}\BibitemShut {NoStop}%
\bibitem [{\citenamefont {Lynch}(1988)}]{Lynch88}%
  \BibitemOpen
  \bibfield  {author} {\bibinfo {author} {\bibfnamefont {S.~P.}\ \bibnamefont
  {Lynch}},\ }\href@noop {} {\bibfield  {journal} {\bibinfo  {journal} {Acta
  Metallurgica},\ }\textbf {\bibinfo {volume} {36}},\ \bibinfo {pages} {2639}
  (\bibinfo {year} {1988})}\BibitemShut {NoStop}%
\bibitem [{\citenamefont {Paxton}\ and\ \citenamefont
  {Finnis}(2008)}]{Paxton08}%
  \BibitemOpen
  \bibfield  {author} {\bibinfo {author} {\bibfnamefont {A.~T.}\ \bibnamefont
  {Paxton}}\ and\ \bibinfo {author} {\bibfnamefont {M.~W.}\ \bibnamefont
  {Finnis}},\ }\href@noop {} {\bibfield  {journal} {\bibinfo  {journal} {Phys.
  Rev. B},\ }\textbf {\bibinfo {volume} {77}},\ \bibinfo {eid} {024428}
  (\bibinfo {year} {2008})}\BibitemShut {NoStop}%
\bibitem [{\citenamefont {Harrison}(1980)}]{Harrison80}%
  \BibitemOpen
  \bibfield  {author} {\bibinfo {author} {\bibfnamefont {W.~A.}\ \bibnamefont
  {Harrison}},\ }\href@noop {} {\emph {\bibinfo {title} {Electronic structure
  and the properties of solids}}}\ (\bibinfo  {publisher} {W.~H.~Freeman},\
  \bibinfo {address} {San Francisco},\ \bibinfo {year} {1980})\BibitemShut
  {NoStop}%
\bibitem [{\citenamefont {Pettifor}(1995)}]{Pettifor95}%
  \BibitemOpen
  \bibfield  {author} {\bibinfo {author} {\bibfnamefont {D.~G.}\ \bibnamefont
  {Pettifor}},\ }\href@noop {} {\emph {\bibinfo {title} {Bonding and structure
  of molecules and solids}}}\ (\bibinfo  {publisher} {Oxford University
  Press},\ \bibinfo {address} {Oxford},\ \bibinfo {year} {1995})\BibitemShut
  {NoStop}%
\bibitem [{\citenamefont {Finnis}(2003)}]{Finnis03}%
  \BibitemOpen
  \bibfield  {author} {\bibinfo {author} {\bibfnamefont {M.~W.}\ \bibnamefont
  {Finnis}},\ }\href@noop {} {\emph {\bibinfo {title} {Interatomic forces in
  condensed matter}}}\ (\bibinfo  {publisher} {Oxford University Press},\
  \bibinfo {address} {Oxford},\ \bibinfo {year} {2003})\BibitemShut {NoStop}%
\bibitem [{\citenamefont {Liu}\ \emph {et~al.}(2005)\citenamefont {Liu},
  \citenamefont {Nguyen-Manh}, \citenamefont {Liu},\ and\ \citenamefont
  {Pettifor}}]{Liu05}%
  \BibitemOpen
  \bibfield  {author} {\bibinfo {author} {\bibfnamefont {G.}~\bibnamefont
  {Liu}}, \bibinfo {author} {\bibfnamefont {D.}~\bibnamefont {Nguyen-Manh}},
  \bibinfo {author} {\bibfnamefont {B.-G.}\ \bibnamefont {Liu}}, \ and\
  \bibinfo {author} {\bibfnamefont {D.~G.}\ \bibnamefont {Pettifor}},\
  }\href@noop {} {\bibfield  {journal} {\bibinfo  {journal} {Phys. Rev. B},\
  }\textbf {\bibinfo {volume} {71}},\ \bibinfo {pages} {174115} (\bibinfo
  {year} {2005})}\BibitemShut {NoStop}%
\bibitem [{Pax()}]{Paxton09}%
  \BibitemOpen
  \href@noop {} {}\bibinfo {note} {A.~T.~Paxton, in {``}Multiscale Simulation
  Methods in Molecular Sciences,{''} (NIC series, vol 42, J{\"u}lich
  Supercomputing Centre) pp.~145--174. Available on-line at {\tt
  http://www.fz-juelich.de/nic-series/volume42}}\BibitemShut {NoStop}%
\bibitem [{\citenamefont {Foulkes}\ and\ \citenamefont
  {Haydock}(1989)}]{Foulkes89}%
  \BibitemOpen
  \bibfield  {author} {\bibinfo {author} {\bibfnamefont {W.~M.~C.}\
  \bibnamefont {Foulkes}}\ and\ \bibinfo {author} {\bibfnamefont
  {R.}~\bibnamefont {Haydock}},\ }\href@noop {} {\bibfield  {journal} {\bibinfo
   {journal} {Phys. Rev. B},\ }\textbf {\bibinfo {volume} {39}},\ \bibinfo
  {pages} {12520} (\bibinfo {year} {1989})}\BibitemShut {NoStop}%
\bibitem [{\citenamefont {Sutton}\ \emph {et~al.}(1988)\citenamefont {Sutton},
  \citenamefont {Finnis}, \citenamefont {Pettifor},\ and\ \citenamefont
  {Ohta}}]{Sutton88}%
  \BibitemOpen
  \bibfield  {author} {\bibinfo {author} {\bibfnamefont {A.~P.}\ \bibnamefont
  {Sutton}}, \bibinfo {author} {\bibfnamefont {M.~W.}\ \bibnamefont {Finnis}},
  \bibinfo {author} {\bibfnamefont {D.~G.}\ \bibnamefont {Pettifor}}, \ and\
  \bibinfo {author} {\bibfnamefont {Y.}~\bibnamefont {Ohta}},\ }\href@noop {}
  {\bibfield  {journal} {\bibinfo  {journal} {Journal of Physics: Condensed
  Matter},\ }\textbf {\bibinfo {volume} {21}},\ \bibinfo {pages} {35} (\bibinfo
  {year} {1988})}\BibitemShut {NoStop}%
\bibitem [{\citenamefont {Slater}\ and\ \citenamefont
  {Koster}(1954)}]{Slater54}%
  \BibitemOpen
  \bibfield  {author} {\bibinfo {author} {\bibfnamefont {J.~C.}\ \bibnamefont
  {Slater}}\ and\ \bibinfo {author} {\bibfnamefont {G.~F.}\ \bibnamefont
  {Koster}},\ }\href@noop {} {\bibfield  {journal} {\bibinfo  {journal} {Phys.
  Rev.},\ }\textbf {\bibinfo {volume} {94}},\ \bibinfo {pages} {1498} (\bibinfo
  {year} {1954})}\BibitemShut {NoStop}%
\bibitem [{\citenamefont {Ducastelle}(1970)}]{Ducastelle70}%
  \BibitemOpen
  \bibfield  {author} {\bibinfo {author} {\bibfnamefont {F.}~\bibnamefont
  {Ducastelle}},\ }\href@noop {} {\bibfield  {journal} {\bibinfo  {journal} {J.
  Phys. France},\ }\textbf {\bibinfo {volume} {31}},\ \bibinfo {pages} {1055}
  (\bibinfo {year} {1970})}\BibitemShut {NoStop}%
\bibitem [{\citenamefont {Spanjaard}\ and\ \citenamefont
  {Desjonqu\`eres}(1984)}]{Spanjaard84}%
  \BibitemOpen
  \bibfield  {author} {\bibinfo {author} {\bibfnamefont {D.}~\bibnamefont
  {Spanjaard}}\ and\ \bibinfo {author} {\bibfnamefont {M.~C.}\ \bibnamefont
  {Desjonqu\`eres}},\ }\href@noop {} {\bibfield  {journal} {\bibinfo  {journal}
  {Phys. Rev. B},\ }\textbf {\bibinfo {volume} {30}},\ \bibinfo {pages} {4822}
  (\bibinfo {year} {1984})}\BibitemShut {NoStop}%
\bibitem [{\citenamefont {Paxton}(1996)}]{Paxton96}%
  \BibitemOpen
  \bibfield  {author} {\bibinfo {author} {\bibfnamefont {A.~T.}\ \bibnamefont
  {Paxton}},\ }\href@noop {} {\bibfield  {journal} {\bibinfo  {journal}
  {Journal of Physics D: Applied Physics},\ }\textbf {\bibinfo {volume} {29}},\
  \bibinfo {pages} {1689} (\bibinfo {year} {1996})}\BibitemShut {NoStop}%
\bibitem [{\citenamefont {Heine}(1967)}]{Heine67}%
  \BibitemOpen
  \bibfield  {author} {\bibinfo {author} {\bibfnamefont {V.}~\bibnamefont
  {Heine}},\ }\href@noop {} {\bibfield  {journal} {\bibinfo  {journal} {Phys.
  Rev.},\ }\textbf {\bibinfo {volume} {153}},\ \bibinfo {pages} {673} (\bibinfo
  {year} {1967})}\BibitemShut {NoStop}%
\bibitem [{\citenamefont {Andersen}(1973)}]{Andersen73}%
  \BibitemOpen
  \bibfield  {author} {\bibinfo {author} {\bibfnamefont {O.~K.}\ \bibnamefont
  {Andersen}},\ }\href@noop {} {\bibfield  {journal} {\bibinfo  {journal}
  {Solid State Communications},\ }\textbf {\bibinfo {volume} {13}},\ \bibinfo
  {pages} {133 } (\bibinfo {year} {1973})}\BibitemShut {NoStop}%
\bibitem [{\citenamefont {Andersen}(1975)}]{Andersen75}%
  \BibitemOpen
  \bibfield  {author} {\bibinfo {author} {\bibfnamefont {O.~K.}\ \bibnamefont
  {Andersen}},\ }\href@noop {} {\bibfield  {journal} {\bibinfo  {journal}
  {Phys. Rev. B},\ }\textbf {\bibinfo {volume} {12}},\ \bibinfo {pages} {3060}
  (\bibinfo {year} {1975})}\BibitemShut {NoStop}%
\bibitem [{\citenamefont {Pettifor}(1977)}]{Pettifor77}%
  \BibitemOpen
  \bibfield  {author} {\bibinfo {author} {\bibfnamefont {D.~G.}\ \bibnamefont
  {Pettifor}},\ }\href@noop {} {\bibfield  {journal} {\bibinfo  {journal}
  {Journal of Physics F: Metal Physics},\ }\textbf {\bibinfo {volume} {7}},\
  \bibinfo {pages} {613} (\bibinfo {year} {1977})}\BibitemShut {NoStop}%
\bibitem [{\citenamefont {Skinner}\ and\ \citenamefont
  {Pettifor}(1991)}]{Skinner91}%
  \BibitemOpen
  \bibfield  {author} {\bibinfo {author} {\bibfnamefont {A.~J.}\ \bibnamefont
  {Skinner}}\ and\ \bibinfo {author} {\bibfnamefont {D.~G.}\ \bibnamefont
  {Pettifor}},\ }\href@noop {} {\bibfield  {journal} {\bibinfo  {journal}
  {Journal of Physics: Condensed Matter},\ }\textbf {\bibinfo {volume} {3}},\
  \bibinfo {pages} {2029} (\bibinfo {year} {1991})}\BibitemShut {NoStop}%
\bibitem [{\citenamefont {Froyen}\ and\ \citenamefont
  {Harrison}(1979)}]{Froyen79}%
  \BibitemOpen
  \bibfield  {author} {\bibinfo {author} {\bibfnamefont {S.}~\bibnamefont
  {Froyen}}\ and\ \bibinfo {author} {\bibfnamefont {W.~A.}\ \bibnamefont
  {Harrison}},\ }\href@noop {} {\bibfield  {journal} {\bibinfo  {journal}
  {Phys. Rev. B},\ }\textbf {\bibinfo {volume} {20}},\ \bibinfo {pages} {2420}
  (\bibinfo {year} {1979})}\BibitemShut {NoStop}%
\bibitem [{\citenamefont {Paxton}\ \emph {et~al.}(1987)\citenamefont {Paxton},
  \citenamefont {Sutton},\ and\ \citenamefont {Nex}}]{Paxton87}%
  \BibitemOpen
  \bibfield  {author} {\bibinfo {author} {\bibfnamefont {A.~T.}\ \bibnamefont
  {Paxton}}, \bibinfo {author} {\bibfnamefont {A.~P.}\ \bibnamefont {Sutton}},
  \ and\ \bibinfo {author} {\bibfnamefont {C.~M.~M.}\ \bibnamefont {Nex}},\
  }\href@noop {} {\bibfield  {journal} {\bibinfo  {journal} {Journal of Physics
  C: Solid State Physics},\ }\textbf {\bibinfo {volume} {20}},\ \bibinfo
  {pages} {L263} (\bibinfo {year} {1987})}\BibitemShut {NoStop}%
\bibitem [{\citenamefont {Goodwin}\ \emph {et~al.}(1989)\citenamefont
  {Goodwin}, \citenamefont {Skinner},\ and\ \citenamefont
  {Pettifor}}]{Goodwin89}%
  \BibitemOpen
  \bibfield  {author} {\bibinfo {author} {\bibfnamefont {L.}~\bibnamefont
  {Goodwin}}, \bibinfo {author} {\bibfnamefont {A.~J.}\ \bibnamefont
  {Skinner}}, \ and\ \bibinfo {author} {\bibfnamefont {D.~G.}\ \bibnamefont
  {Pettifor}},\ }\href@noop {} {\bibfield  {journal} {\bibinfo  {journal}
  {Europhys.~Lett.},\ }\textbf {\bibinfo {volume} {9}},\ \bibinfo {pages} {701}
  (\bibinfo {year} {1989})}\BibitemShut {NoStop}%
\bibitem [{\citenamefont {Znam}\ \emph {et~al.}(2003)\citenamefont {Znam},
  \citenamefont {Nguyen-{M}anh}, \citenamefont {Pettifor},\ and\ \citenamefont
  {Vitek}}]{Znam03}%
  \BibitemOpen
  \bibfield  {author} {\bibinfo {author} {\bibfnamefont {S.}~\bibnamefont
  {Znam}}, \bibinfo {author} {\bibfnamefont {D.}~\bibnamefont {Nguyen-{M}anh}},
  \bibinfo {author} {\bibfnamefont {D.~G.}\ \bibnamefont {Pettifor}}, \ and\
  \bibinfo {author} {\bibfnamefont {V.}~\bibnamefont {Vitek}},\ }\href@noop {}
  {\bibfield  {journal} {\bibinfo  {journal} {Phil. Mag. A},\ }\textbf
  {\bibinfo {volume} {83}},\ \bibinfo {pages} {415} (\bibinfo {year}
  {2003})}\BibitemShut {NoStop}%
\bibitem [{\citenamefont {Allen}\ \emph {et~al.}(1986)\citenamefont {Allen},
  \citenamefont {Broughton},\ and\ \citenamefont {McMahan}}]{Allen86}%
  \BibitemOpen
  \bibfield  {author} {\bibinfo {author} {\bibfnamefont {P.~B.}\ \bibnamefont
  {Allen}}, \bibinfo {author} {\bibfnamefont {J.~Q.}\ \bibnamefont
  {Broughton}}, \ and\ \bibinfo {author} {\bibfnamefont {A.~K.}\ \bibnamefont
  {McMahan}},\ }\href@noop {} {\bibfield  {journal} {\bibinfo  {journal} {Phys.
  Rev. B},\ }\textbf {\bibinfo {volume} {34}},\ \bibinfo {pages} {859}
  (\bibinfo {year} {1986})}\BibitemShut {NoStop}%
\bibitem [{\citenamefont {Pettifor}(1972)}]{Pettifor72}%
  \BibitemOpen
  \bibfield  {author} {\bibinfo {author} {\bibfnamefont {D.~G.}\ \bibnamefont
  {Pettifor}},\ }\href@noop {} {\bibfield  {journal} {\bibinfo  {journal}
  {Journal of Physics F: Metal Physics},\ }\textbf {\bibinfo {volume} {5}},\
  \bibinfo {pages} {97} (\bibinfo {year} {1972})}\BibitemShut {NoStop}%
\bibitem [{\citenamefont {Andersen}\ and\ \citenamefont
  {Jepsen}(1984)}]{Andersen84}%
  \BibitemOpen
  \bibfield  {author} {\bibinfo {author} {\bibfnamefont {O.~K.}\ \bibnamefont
  {Andersen}}\ and\ \bibinfo {author} {\bibfnamefont {O.}~\bibnamefont
  {Jepsen}},\ }\href@noop {} {\bibfield  {journal} {\bibinfo  {journal} {Phys.
  Rev. Lett.},\ }\textbf {\bibinfo {volume} {53}},\ \bibinfo {pages} {2571}
  (\bibinfo {year} {1984})}\BibitemShut {NoStop}%
\bibitem [{\citenamefont {Tang}\ \emph {et~al.}(1996)\citenamefont {Tang},
  \citenamefont {Wang}, \citenamefont {Chan},\ and\ \citenamefont
  {Ho}}]{Tang96}%
  \BibitemOpen
  \bibfield  {author} {\bibinfo {author} {\bibfnamefont {M.~S.}\ \bibnamefont
  {Tang}}, \bibinfo {author} {\bibfnamefont {C.~Z.}\ \bibnamefont {Wang}},
  \bibinfo {author} {\bibfnamefont {C.~T.}\ \bibnamefont {Chan}}, \ and\
  \bibinfo {author} {\bibfnamefont {K.~M.}\ \bibnamefont {Ho}},\ }\href@noop {}
  {\bibfield  {journal} {\bibinfo  {journal} {Phys. Rev. B},\ }\textbf
  {\bibinfo {volume} {53}},\ \bibinfo {pages} {979} (\bibinfo {year}
  {1996})}\BibitemShut {NoStop}%
\bibitem [{\citenamefont {Haas}\ \emph {et~al.}(1998)\citenamefont {Haas},
  \citenamefont {Wang}, \citenamefont {F\"ahnle}, \citenamefont {Els\"asser},\
  and\ \citenamefont {Ho}}]{Haas98}%
  \BibitemOpen
  \bibfield  {author} {\bibinfo {author} {\bibfnamefont {H.}~\bibnamefont
  {Haas}}, \bibinfo {author} {\bibfnamefont {C.~Z.}\ \bibnamefont {Wang}},
  \bibinfo {author} {\bibfnamefont {M.}~\bibnamefont {F\"ahnle}}, \bibinfo
  {author} {\bibfnamefont {C.}~\bibnamefont {Els\"asser}}, \ and\ \bibinfo
  {author} {\bibfnamefont {K.~M.}\ \bibnamefont {Ho}},\ }\href@noop {}
  {\bibfield  {journal} {\bibinfo  {journal} {Phys. Rev. B},\ }\textbf
  {\bibinfo {volume} {57}},\ \bibinfo {pages} {1461} (\bibinfo {year}
  {1998})}\BibitemShut {NoStop}%
\bibitem [{\citenamefont {Nguyen-Manh}\ \emph {et~al.}(2000)\citenamefont
  {Nguyen-Manh}, \citenamefont {Pettifor},\ and\ \citenamefont
  {Vitek}}]{Pettifor00}%
  \BibitemOpen
  \bibfield  {author} {\bibinfo {author} {\bibfnamefont {D.}~\bibnamefont
  {Nguyen-Manh}}, \bibinfo {author} {\bibfnamefont {D.~G.}\ \bibnamefont
  {Pettifor}}, \ and\ \bibinfo {author} {\bibfnamefont {V.}~\bibnamefont
  {Vitek}},\ }\href@noop {} {\bibfield  {journal} {\bibinfo  {journal} {Phys.
  Rev. Lett.},\ }\textbf {\bibinfo {volume} {85}},\ \bibinfo {pages} {4136}
  (\bibinfo {year} {2000})}\BibitemShut {NoStop}%
\bibitem [{\citenamefont {Mrovec}\ \emph {et~al.}(2004)\citenamefont {Mrovec},
  \citenamefont {Nguyen-Manh}, \citenamefont {Pettifor},\ and\ \citenamefont
  {Vitek}}]{Mrovec04}%
  \BibitemOpen
  \bibfield  {author} {\bibinfo {author} {\bibfnamefont {M.}~\bibnamefont
  {Mrovec}}, \bibinfo {author} {\bibfnamefont {D.}~\bibnamefont {Nguyen-Manh}},
  \bibinfo {author} {\bibfnamefont {D.~G.}\ \bibnamefont {Pettifor}}, \ and\
  \bibinfo {author} {\bibfnamefont {V.}~\bibnamefont {Vitek}},\ }\href@noop {}
  {\bibfield  {journal} {\bibinfo  {journal} {Phys. Rev. B},\ }\textbf
  {\bibinfo {volume} {69}},\ \bibinfo {pages} {094115} (\bibinfo {year}
  {2004})}\BibitemShut {NoStop}%
\bibitem [{\citenamefont {Mrovec}\ \emph {et~al.}(2007)\citenamefont {Mrovec},
  \citenamefont {Gr\"{o}ger}, \citenamefont {Bailey}, \citenamefont
  {Nguyen-Manh}, \citenamefont {Els\"{a}sser},\ and\ \citenamefont
  {Vitek}}]{Mrovec07}%
  \BibitemOpen
  \bibfield  {author} {\bibinfo {author} {\bibfnamefont {M.}~\bibnamefont
  {Mrovec}}, \bibinfo {author} {\bibfnamefont {R.}~\bibnamefont {Gr\"{o}ger}},
  \bibinfo {author} {\bibfnamefont {A.~G.}\ \bibnamefont {Bailey}}, \bibinfo
  {author} {\bibfnamefont {D.}~\bibnamefont {Nguyen-Manh}}, \bibinfo {author}
  {\bibfnamefont {C.}~\bibnamefont {Els\"{a}sser}}, \ and\ \bibinfo {author}
  {\bibfnamefont {V.}~\bibnamefont {Vitek}},\ }\href@noop {} {\bibfield
  {journal} {\bibinfo  {journal} {Phys. Rev. B},\ }\textbf {\bibinfo {volume}
  {75}},\ \bibinfo {eid} {104119} (\bibinfo {year} {2007})}\BibitemShut
  {NoStop}%
\bibitem [{\citenamefont {Perdew}\ \emph {et~al.}(1997)\citenamefont {Perdew},
  \citenamefont {Burke},\ and\ \citenamefont {Ernzerhof}}]{PBE}%
  \BibitemOpen
  \bibfield  {author} {\bibinfo {author} {\bibfnamefont {J.~P.}\ \bibnamefont
  {Perdew}}, \bibinfo {author} {\bibfnamefont {K.}~\bibnamefont {Burke}}, \
  and\ \bibinfo {author} {\bibfnamefont {M.}~\bibnamefont {Ernzerhof}},\
  }\href@noop {} {\bibfield  {journal} {\bibinfo  {journal} {Phys. Rev.
  Lett.},\ }\textbf {\bibinfo {volume} {78}},\ \bibinfo {pages} {1396}
  (\bibinfo {year} {1997})}\BibitemShut {NoStop}%
\bibitem [{\citenamefont {Methfessel}\ \emph {et~al.}(2000)\citenamefont
  {Methfessel}, \citenamefont {van Schilfgaarde},\ and\ \citenamefont
  {Casali}}]{Methfessel00}%
  \BibitemOpen
  \bibfield  {author} {\bibinfo {author} {\bibfnamefont {M.}~\bibnamefont
  {Methfessel}}, \bibinfo {author} {\bibfnamefont {M.}~\bibnamefont {van
  Schilfgaarde}}, \ and\ \bibinfo {author} {\bibfnamefont {R.~A.}\ \bibnamefont
  {Casali}},\ }\enquote {\bibinfo {title} {Electronic structure and physical
  properties of solids: the uses of the {LMTO} method},}\ \ (\bibinfo
  {publisher} {Springer-Verlag},\ \bibinfo {address} {Berlin},\ \bibinfo {year}
  {2000})\ pp.\ \bibinfo {pages} {114--147}\BibitemShut {NoStop}%
\bibitem [{\citenamefont {Pettifor}(1979)}]{Pettifor79}%
  \BibitemOpen
  \bibfield  {author} {\bibinfo {author} {\bibfnamefont {D.~G.}\ \bibnamefont
  {Pettifor}},\ }\href@noop {} {\bibfield  {journal} {\bibinfo  {journal}
  {Phys. Rev. Lett.},\ }\textbf {\bibinfo {volume} {42}},\ \bibinfo {pages}
  {846} (\bibinfo {year} {1979})}\BibitemShut {NoStop}%
\bibitem [{\citenamefont {Williams}\ \emph {et~al.}(1980)\citenamefont
  {Williams}, \citenamefont {Gelatt},\ and\ \citenamefont
  {Moruzzi}}]{Williams80}%
  \BibitemOpen
  \bibfield  {author} {\bibinfo {author} {\bibfnamefont {A.~R.}\ \bibnamefont
  {Williams}}, \bibinfo {author} {\bibfnamefont {C.~D.}\ \bibnamefont
  {Gelatt}}, \ and\ \bibinfo {author} {\bibfnamefont {V.~L.}\ \bibnamefont
  {Moruzzi}},\ }\href@noop {} {\bibfield  {journal} {\bibinfo  {journal} {Phys.
  Rev. Lett.},\ }\textbf {\bibinfo {volume} {44}},\ \bibinfo {pages} {429}
  (\bibinfo {year} {1980})}\BibitemShut {NoStop}%
\bibitem [{\citenamefont {Pettifor}(1987)}]{Pettifor87}%
  \BibitemOpen
  \bibfield  {author} {\bibinfo {author} {\bibfnamefont {D.~G.}\ \bibnamefont
  {Pettifor}},\ }\href@noop {} {\bibfield  {journal} {\bibinfo  {journal}
  {Solid State Physics},\ }\textbf {\bibinfo {volume} {40}},\ \bibinfo {pages}
  {43} (\bibinfo {year} {1987})}\BibitemShut {NoStop}%
\bibitem [{\citenamefont {Philipsen}\ and\ \citenamefont
  {Baerends}(1996)}]{Philipsen96}%
  \BibitemOpen
  \bibfield  {author} {\bibinfo {author} {\bibfnamefont {P.~H.~T.}\
  \bibnamefont {Philipsen}}\ and\ \bibinfo {author} {\bibfnamefont {E.~J.}\
  \bibnamefont {Baerends}},\ }\href@noop {} {\bibfield  {journal} {\bibinfo
  {journal} {Phys. Rev. B},\ }\textbf {\bibinfo {volume} {54}},\ \bibinfo
  {pages} {5326} (\bibinfo {year} {1996})}\BibitemShut {NoStop}%
\bibitem [{\citenamefont {Seki}\ and\ \citenamefont {Nagata}(2005)}]{Seki05}%
  \BibitemOpen
  \bibfield  {author} {\bibinfo {author} {\bibfnamefont {I.}~\bibnamefont
  {Seki}}\ and\ \bibinfo {author} {\bibfnamefont {K.}~\bibnamefont {Nagata}},\
  }\href@noop {} {\bibfield  {journal} {\bibinfo  {journal} {ISIJ
  International},\ }\textbf {\bibinfo {volume} {45}},\ \bibinfo {pages} {1789}
  (\bibinfo {year} {2005})}\BibitemShut {NoStop}%
\bibitem [{\citenamefont {Zarestky}\ and\ \citenamefont
  {Stassis}(1987)}]{Zarestky87}%
  \BibitemOpen
  \bibfield  {author} {\bibinfo {author} {\bibfnamefont {J.}~\bibnamefont
  {Zarestky}}\ and\ \bibinfo {author} {\bibfnamefont {C.}~\bibnamefont
  {Stassis}},\ }\href@noop {} {\bibfield  {journal} {\bibinfo  {journal} {Phys.
  Rev. B},\ }\textbf {\bibinfo {volume} {35}},\ \bibinfo {pages} {4500}
  (\bibinfo {year} {1987})}\BibitemShut {NoStop}%
\bibitem [{\citenamefont {Edwards}(1983)}]{Edwards83}%
  \BibitemOpen
  \bibfield  {author} {\bibinfo {author} {\bibfnamefont {D.~M.}\ \bibnamefont
  {Edwards}},\ }\href@noop {} {\bibfield  {journal} {\bibinfo  {journal}
  {Journal of Magnetism and Magnetic Materials},\ }\textbf {\bibinfo {volume}
  {36}},\ \bibinfo {pages} {213} (\bibinfo {year} {1983})}\BibitemShut
  {NoStop}%
\bibitem [{\citenamefont {Adams}\ \emph {et~al.}(2006)\citenamefont {Adams},
  \citenamefont {Agosta}, \citenamefont {Leisure},\ and\ \citenamefont
  {Ledbetter}}]{Adams06}%
  \BibitemOpen
  \bibfield  {author} {\bibinfo {author} {\bibfnamefont {J.~J.}\ \bibnamefont
  {Adams}}, \bibinfo {author} {\bibfnamefont {D.~S.}\ \bibnamefont {Agosta}},
  \bibinfo {author} {\bibfnamefont {R.~G.}\ \bibnamefont {Leisure}}, \ and\
  \bibinfo {author} {\bibfnamefont {H.}~\bibnamefont {Ledbetter}},\ }\href@noop
  {} {\bibfield  {journal} {\bibinfo  {journal} {J. Appl. Phys.},\ }\textbf
  {\bibinfo {volume} {100}},\ \bibinfo {pages} {113530} (\bibinfo {year}
  {2006})}\BibitemShut {NoStop}%
\bibitem [{\citenamefont {Guo}\ and\ \citenamefont {Wang}(2000)}]{Guo00}%
  \BibitemOpen
  \bibfield  {author} {\bibinfo {author} {\bibfnamefont {G.~Y.}\ \bibnamefont
  {Guo}}\ and\ \bibinfo {author} {\bibfnamefont {H.~H.}\ \bibnamefont {Wang}},\
  }\href@noop {} {\bibfield  {journal} {\bibinfo  {journal} {Chinese J.
  Phys.},\ }\textbf {\bibinfo {volume} {38}},\ \bibinfo {pages} {949} (\bibinfo
  {year} {2000})}\BibitemShut {NoStop}%
\bibitem [{\citenamefont {Heine}\ \emph {et~al.}(1980)\citenamefont {Heine},
  \citenamefont {Holden}, \citenamefont {Lin-{C}hung},\ and\ \citenamefont
  {You}}]{Heine80}%
  \BibitemOpen
  \bibfield  {author} {\bibinfo {author} {\bibfnamefont {V.}~\bibnamefont
  {Heine}}, \bibinfo {author} {\bibfnamefont {A.}~\bibnamefont {Holden}},
  \bibinfo {author} {\bibfnamefont {P.}~\bibnamefont {Lin-{C}hung}}, \ and\
  \bibinfo {author} {\bibfnamefont {M.}~\bibnamefont {You}},\ }\href@noop {}
  {\bibfield  {journal} {\bibinfo  {journal} {Journal of Magnetism and Magnetic
  Materials},\ }\textbf {\bibinfo {volume} {15-18}},\ \bibinfo {pages} {69}
  (\bibinfo {year} {1980})}\BibitemShut {NoStop}%
\bibitem [{\citenamefont {Hasegawa}\ and\ \citenamefont
  {Pettifor}(1983)}]{Pettifor83}%
  \BibitemOpen
  \bibfield  {author} {\bibinfo {author} {\bibfnamefont {H.}~\bibnamefont
  {Hasegawa}}\ and\ \bibinfo {author} {\bibfnamefont {D.~G.}\ \bibnamefont
  {Pettifor}},\ }\href@noop {} {\bibfield  {journal} {\bibinfo  {journal}
  {Phys. Rev. Lett.},\ }\textbf {\bibinfo {volume} {50}},\ \bibinfo {pages}
  {130} (\bibinfo {year} {1983})}\BibitemShut {NoStop}%
\bibitem [{\citenamefont {Leslie}(1981)}]{Leslie}%
  \BibitemOpen
  \bibfield  {author} {\bibinfo {author} {\bibfnamefont {W.~C.}\ \bibnamefont
  {Leslie}},\ }\href@noop {} {\emph {\bibinfo {title} {The Physical Metallurgy
  of Steels}}}\ (\bibinfo  {publisher} {Hemisphere},\ \bibinfo {address}
  {Washington},\ \bibinfo {year} {1981})\BibitemShut {NoStop}%
\bibitem [{\citenamefont {Kaufman}\ \emph {et~al.}(1963)\citenamefont
  {Kaufman}, \citenamefont {Clougherty},\ and\ \citenamefont
  {Weiss}}]{Kaufman63}%
  \BibitemOpen
  \bibfield  {author} {\bibinfo {author} {\bibfnamefont {L.}~\bibnamefont
  {Kaufman}}, \bibinfo {author} {\bibfnamefont {E.}~\bibnamefont {Clougherty}},
  \ and\ \bibinfo {author} {\bibfnamefont {R.~J.}\ \bibnamefont {Weiss}},\
  }\href@noop {} {\bibfield  {journal} {\bibinfo  {journal} {Acta
  Metallurgica},\ }\textbf {\bibinfo {volume} {11}},\ \bibinfo {pages} {323}
  (\bibinfo {year} {1963})}\BibitemShut {NoStop}%
\bibitem [{\citenamefont {Roy}\ and\ \citenamefont {Pettifor}(1977)}]{Roy77}%
  \BibitemOpen
  \bibfield  {author} {\bibinfo {author} {\bibfnamefont {D.~M.}\ \bibnamefont
  {Roy}}\ and\ \bibinfo {author} {\bibfnamefont {D.~G.}\ \bibnamefont
  {Pettifor}},\ }\href@noop {} {\bibfield  {journal} {\bibinfo  {journal}
  {J.~Phys.~F: Metal Phys.},\ }\textbf {\bibinfo {volume} {7}},\ \bibinfo
  {pages} {L183} (\bibinfo {year} {1977})}\BibitemShut {NoStop}%
\bibitem [{\citenamefont {Christensen}\ \emph {et~al.}(1988)\citenamefont
  {Christensen}, \citenamefont {Gunnarsson}, \citenamefont {Jepsen},\ and\
  \citenamefont {Andersen}}]{Christensen88}%
  \BibitemOpen
  \bibfield  {author} {\bibinfo {author} {\bibfnamefont {N.~E.}\ \bibnamefont
  {Christensen}}, \bibinfo {author} {\bibfnamefont {O.}~\bibnamefont
  {Gunnarsson}}, \bibinfo {author} {\bibfnamefont {O.}~\bibnamefont {Jepsen}},
  \ and\ \bibinfo {author} {\bibfnamefont {O.~K.}\ \bibnamefont {Andersen}},\
  }\href@noop {} {\bibfield  {journal} {\bibinfo  {journal}
  {J.~Phys.~Colloques~C8},\ }\textbf {\bibinfo {volume} {49}},\ \bibinfo
  {pages} {17} (\bibinfo {year} {1988})}\BibitemShut {NoStop}%
\bibitem [{\citenamefont {Els\"asser}\ \emph
  {et~al.}(1998){\natexlab{a}}\citenamefont {Els\"asser}, \citenamefont {Zhu},
  \citenamefont {Louie}, \citenamefont {F\"ahnle},\ and\ \citenamefont
  {Chan}}]{CE1}%
  \BibitemOpen
  \bibfield  {author} {\bibinfo {author} {\bibfnamefont {C.}~\bibnamefont
  {Els\"asser}}, \bibinfo {author} {\bibfnamefont {J.}~\bibnamefont {Zhu}},
  \bibinfo {author} {\bibfnamefont {S.~G.}\ \bibnamefont {Louie}}, \bibinfo
  {author} {\bibfnamefont {M.}~\bibnamefont {F\"ahnle}}, \ and\ \bibinfo
  {author} {\bibfnamefont {C.~T.}\ \bibnamefont {Chan}},\ }\href@noop {}
  {\bibfield  {journal} {\bibinfo  {journal} {Journal of Physics: Condensed
  Matter},\ }\textbf {\bibinfo {volume} {10}},\ \bibinfo {pages} {5081}
  (\bibinfo {year} {1998}{\natexlab{a}})}\BibitemShut {NoStop}%
\bibitem [{\citenamefont {Tajima}\ \emph {et~al.}(1976)\citenamefont {Tajima},
  \citenamefont {Endoh}, \citenamefont {Ishikawa},\ and\ \citenamefont
  {Stirling}}]{Tajima76}%
  \BibitemOpen
  \bibfield  {author} {\bibinfo {author} {\bibfnamefont {K.}~\bibnamefont
  {Tajima}}, \bibinfo {author} {\bibfnamefont {Y.}~\bibnamefont {Endoh}},
  \bibinfo {author} {\bibfnamefont {Y.}~\bibnamefont {Ishikawa}}, \ and\
  \bibinfo {author} {\bibfnamefont {W.~G.}\ \bibnamefont {Stirling}},\
  }\href@noop {} {\bibfield  {journal} {\bibinfo  {journal} {Phys. Rev.
  Lett.},\ }\textbf {\bibinfo {volume} {37}},\ \bibinfo {pages} {519} (\bibinfo
  {year} {1976})}\BibitemShut {NoStop}%
\bibitem [{\citenamefont {Hasegawa}\ \emph {et~al.}(1985)\citenamefont
  {Hasegawa}, \citenamefont {Finnis},\ and\ \citenamefont
  {Pettifor}}]{Hasegawa85}%
  \BibitemOpen
  \bibfield  {author} {\bibinfo {author} {\bibfnamefont {H.}~\bibnamefont
  {Hasegawa}}, \bibinfo {author} {\bibfnamefont {M.~W.}\ \bibnamefont
  {Finnis}}, \ and\ \bibinfo {author} {\bibfnamefont {D.~G.}\ \bibnamefont
  {Pettifor}},\ }\href@noop {} {\bibfield  {journal} {\bibinfo  {journal}
  {Journal of Physics F: Metal Physics},\ }\textbf {\bibinfo {volume} {15}},\
  \bibinfo {pages} {19} (\bibinfo {year} {1985})}\BibitemShut {NoStop}%
\bibitem [{\citenamefont {Hasegawa}\ \emph {et~al.}(1987)\citenamefont
  {Hasegawa}, \citenamefont {Finnis},\ and\ \citenamefont
  {Pettifor}}]{Hasegawa87}%
  \BibitemOpen
  \bibfield  {author} {\bibinfo {author} {\bibfnamefont {H.}~\bibnamefont
  {Hasegawa}}, \bibinfo {author} {\bibfnamefont {M.~W.}\ \bibnamefont
  {Finnis}}, \ and\ \bibinfo {author} {\bibfnamefont {D.~G.}\ \bibnamefont
  {Pettifor}},\ }\href@noop {} {\bibfield  {journal} {\bibinfo  {journal}
  {Journal of Physics F: Metal Physics},\ }\textbf {\bibinfo {volume} {17}},\
  \bibinfo {pages} {2049} (\bibinfo {year} {1987})}\BibitemShut {NoStop}%
\bibitem [{\citenamefont {Spencer}\ \emph {et~al.}(2002)\citenamefont
  {Spencer}, \citenamefont {Hung}, \citenamefont {Snook},\ and\ \citenamefont
  {Yarovsky}}]{Spencer02}%
  \BibitemOpen
  \bibfield  {author} {\bibinfo {author} {\bibfnamefont {M.~J.~S.}\
  \bibnamefont {Spencer}}, \bibinfo {author} {\bibfnamefont {A.}~\bibnamefont
  {Hung}}, \bibinfo {author} {\bibfnamefont {I.~K.}\ \bibnamefont {Snook}}, \
  and\ \bibinfo {author} {\bibfnamefont {I.}~\bibnamefont {Yarovsky}},\
  }\href@noop {} {\bibfield  {journal} {\bibinfo  {journal} {Surf. Sci.},\
  }\textbf {\bibinfo {volume} {513}},\ \bibinfo {pages} {389} (\bibinfo {year}
  {2002})}\BibitemShut {NoStop}%
\bibitem [{\citenamefont {Finnis}\ \emph
  {et~al.}(1998){\natexlab{a}}\citenamefont {Finnis}, \citenamefont {Paxton},
  \citenamefont {Methfessel},\ and\ \citenamefont {van
  Schilfgaarde}}]{Finnis98b}%
  \BibitemOpen
  \bibfield  {author} {\bibinfo {author} {\bibfnamefont {M.~W.}\ \bibnamefont
  {Finnis}}, \bibinfo {author} {\bibfnamefont {A.~T.}\ \bibnamefont {Paxton}},
  \bibinfo {author} {\bibfnamefont {M.}~\bibnamefont {Methfessel}}, \ and\
  \bibinfo {author} {\bibfnamefont {M.}~\bibnamefont {van Schilfgaarde}},\ }in\
  \href@noop {} {\emph {\bibinfo {booktitle} {Tight binding approach to
  computational materials science, MRS Symp.~Proc.~No.~491}}},\ \bibinfo
  {editor} {edited by\ \bibinfo {editor} {\bibfnamefont {P.~E.~A.}\
  \bibnamefont {Turchi}}, \bibinfo {editor} {\bibfnamefont {A.}~\bibnamefont
  {Gonis}}, \ and\ \bibinfo {editor} {\bibfnamefont {L.}~\bibnamefont
  {Colombo}}}\ (\bibinfo  {publisher} {Materials Research Society},\ \bibinfo
  {address} {Pittsburgh PA},\ \bibinfo {year} {1998})\ pp.\ \bibinfo {pages}
  {265--74}\BibitemShut {NoStop}%
\bibitem [{\citenamefont {Allen}\ \emph {et~al.}(1956)\citenamefont {Allen},
  \citenamefont {Hopkins},\ and\ \citenamefont {Mc{L}ennan}}]{Allen56}%
  \BibitemOpen
  \bibfield  {author} {\bibinfo {author} {\bibfnamefont {N.~P.}\ \bibnamefont
  {Allen}}, \bibinfo {author} {\bibfnamefont {B.~E.}\ \bibnamefont {Hopkins}},
  \ and\ \bibinfo {author} {\bibfnamefont {J.~E.}\ \bibnamefont {Mc{L}ennan}},\
  }\href@noop {} {\bibfield  {journal} {\bibinfo  {journal} {Proc. R. Soc.
  Lond. A},\ }\textbf {\bibinfo {volume} {234}},\ \bibinfo {pages} {221}
  (\bibinfo {year} {1956})}\BibitemShut {NoStop}%
\bibitem [{\citenamefont {Ayer}\ \emph {et~al.}(2006)\citenamefont {Ayer},
  \citenamefont {Mueller},\ and\ \citenamefont {Neeraj}}]{Ayer06}%
  \BibitemOpen
  \bibfield  {author} {\bibinfo {author} {\bibfnamefont {R.}~\bibnamefont
  {Ayer}}, \bibinfo {author} {\bibfnamefont {R.}~\bibnamefont {Mueller}}, \
  and\ \bibinfo {author} {\bibfnamefont {T.}~\bibnamefont {Neeraj}},\
  }\href@noop {} {\bibfield  {journal} {\bibinfo  {journal} {Materials Science
  and Engineering: A},\ }\textbf {\bibinfo {volume} {417}},\ \bibinfo {pages}
  {243} (\bibinfo {year} {2006})}\BibitemShut {NoStop}%
\bibitem [{\citenamefont {Nakasato}\ and\ \citenamefont
  {Bernstein}(1978)}]{Nakasato78}%
  \BibitemOpen
  \bibfield  {author} {\bibinfo {author} {\bibfnamefont {F.}~\bibnamefont
  {Nakasato}}\ and\ \bibinfo {author} {\bibfnamefont {I.~M.}\ \bibnamefont
  {Bernstein}},\ }\href@noop {} {\bibfield  {journal} {\bibinfo  {journal}
  {Metallurgical Transactions A},\ }\textbf {\bibinfo {volume} {9A}},\ \bibinfo
  {pages} {1317} (\bibinfo {year} {1978})}\BibitemShut {NoStop}%
\bibitem [{\citenamefont {Gillan}(1989)}]{Gillan89}%
  \BibitemOpen
  \bibfield  {author} {\bibinfo {author} {\bibfnamefont {M.~J.}\ \bibnamefont
  {Gillan}},\ }\href@noop {} {\bibfield  {journal} {\bibinfo  {journal}
  {Journal of Physics: Condensed Matter},\ }\textbf {\bibinfo {volume} {1}},\
  \bibinfo {pages} {689} (\bibinfo {year} {1989})}\BibitemShut {NoStop}%
\bibitem [{\citenamefont {Domain}\ and\ \citenamefont
  {Becquart}(2001)}]{Domain01}%
  \BibitemOpen
  \bibfield  {author} {\bibinfo {author} {\bibfnamefont {C.}~\bibnamefont
  {Domain}}\ and\ \bibinfo {author} {\bibfnamefont {C.~S.}\ \bibnamefont
  {Becquart}},\ }\href@noop {} {\bibfield  {journal} {\bibinfo  {journal}
  {Phys. Rev. B},\ }\textbf {\bibinfo {volume} {65}},\ \bibinfo {pages}
  {024103} (\bibinfo {year} {2001})}\BibitemShut {NoStop}%
\bibitem [{\citenamefont {S\"oderlind}\ \emph {et~al.}(2000)\citenamefont
  {S\"oderlind}, \citenamefont {Yang}, \citenamefont {Moriarty},\ and\
  \citenamefont {Wills}}]{Soderlind00}%
  \BibitemOpen
  \bibfield  {author} {\bibinfo {author} {\bibfnamefont {P.}~\bibnamefont
  {S\"oderlind}}, \bibinfo {author} {\bibfnamefont {L.~H.}\ \bibnamefont
  {Yang}}, \bibinfo {author} {\bibfnamefont {J.~A.}\ \bibnamefont {Moriarty}},
  \ and\ \bibinfo {author} {\bibfnamefont {J.~M.}\ \bibnamefont {Wills}},\
  }\href@noop {} {\bibfield  {journal} {\bibinfo  {journal} {Phys. Rev. B},\
  }\textbf {\bibinfo {volume} {61}},\ \bibinfo {pages} {2579} (\bibinfo {year}
  {2000})}\BibitemShut {NoStop}%
\bibitem [{\citenamefont {Tateyama}\ and\ \citenamefont
  {Ohno}(2003)}]{Tateyama03}%
  \BibitemOpen
  \bibfield  {author} {\bibinfo {author} {\bibfnamefont {Y.}~\bibnamefont
  {Tateyama}}\ and\ \bibinfo {author} {\bibfnamefont {T.}~\bibnamefont
  {Ohno}},\ }\href@noop {} {\bibfield  {journal} {\bibinfo  {journal} {Phys.
  Rev. B},\ }\textbf {\bibinfo {volume} {67}},\ \bibinfo {pages} {174105}
  (\bibinfo {year} {2003})}\BibitemShut {NoStop}%
\bibitem [{\citenamefont {F{\"u}rderer}\ \emph {et~al.}(1987)\citenamefont
  {F{\"u}rderer}, \citenamefont {D{\"o}ring}, \citenamefont {Gladisch},
  \citenamefont {Haas}, \citenamefont {Herlach}, \citenamefont {Major},
  \citenamefont {Mundinger}, \citenamefont {Rosenkranz}, \citenamefont
  {Sch{\"a}fer}, \citenamefont {Schimmele}, \citenamefont {Schwartz},\ and\
  \citenamefont {Seeger}}]{Furderer87}%
  \BibitemOpen
  \bibfield  {author} {\bibinfo {author} {\bibfnamefont {K.}~\bibnamefont
  {F{\"u}rderer}}, \bibinfo {author} {\bibfnamefont {K.-P.}\ \bibnamefont
  {D{\"o}ring}}, \bibinfo {author} {\bibfnamefont {M.}~\bibnamefont
  {Gladisch}}, \bibinfo {author} {\bibfnamefont {N.}~\bibnamefont {Haas}},
  \bibinfo {author} {\bibfnamefont {D.}~\bibnamefont {Herlach}}, \bibinfo
  {author} {\bibfnamefont {J.}~\bibnamefont {Major}}, \bibinfo {author}
  {\bibfnamefont {H.-J.}\ \bibnamefont {Mundinger}}, \bibinfo {author}
  {\bibfnamefont {J.}~\bibnamefont {Rosenkranz}}, \bibinfo {author}
  {\bibfnamefont {W.}~\bibnamefont {Sch{\"a}fer}}, \bibinfo {author}
  {\bibfnamefont {L.}~\bibnamefont {Schimmele}}, \bibinfo {author}
  {\bibfnamefont {W.}~\bibnamefont {Schwartz}}, \ and\ \bibinfo {author}
  {\bibfnamefont {A.}~\bibnamefont {Seeger}},\ }\href@noop {} {\bibfield
  {journal} {\bibinfo  {journal} {Materials Science Forum},\ }\textbf {\bibinfo
  {volume} {15--18}},\ \bibinfo {pages} {125} (\bibinfo {year}
  {1987})}\BibitemShut {NoStop}%
\bibitem [{\citenamefont {Seeger}(1998)}]{Seeger98}%
  \BibitemOpen
  \bibfield  {author} {\bibinfo {author} {\bibfnamefont {A.}~\bibnamefont
  {Seeger}},\ }\href@noop {} {\bibfield  {journal} {\bibinfo  {journal} {phys.
  stat. sol. (a)},\ }\textbf {\bibinfo {volume} {167}},\ \bibinfo {pages} {289}
  (\bibinfo {year} {1998})}\BibitemShut {NoStop}%
\bibitem [{\citenamefont {Seydel}\ \emph {et~al.}(1994)\citenamefont {Seydel},
  \citenamefont {Frohberg},\ and\ \citenamefont {Wever}}]{Seydel94}%
  \BibitemOpen
  \bibfield  {author} {\bibinfo {author} {\bibfnamefont {O.}~\bibnamefont
  {Seydel}}, \bibinfo {author} {\bibfnamefont {G.}~\bibnamefont {Frohberg}}, \
  and\ \bibinfo {author} {\bibfnamefont {H.}~\bibnamefont {Wever}},\
  }\href@noop {} {\bibfield  {journal} {\bibinfo  {journal} {phys. stat. sol.
  (a)},\ }\textbf {\bibinfo {volume} {144}},\ \bibinfo {pages} {69} (\bibinfo
  {year} {1994})}\BibitemShut {NoStop}%
\bibitem [{\citenamefont {De~{S}chepper}\ \emph {et~al.}(1983)\citenamefont
  {De~{S}chepper}, \citenamefont {Segers}, \citenamefont {Dorikens-Vanpraet},
  \citenamefont {Dorikens}, \citenamefont {Knuyt}, \citenamefont {Stals},\ and\
  \citenamefont {Moser}}]{DeSchepper83}%
  \BibitemOpen
  \bibfield  {author} {\bibinfo {author} {\bibfnamefont {L.}~\bibnamefont
  {De~{S}chepper}}, \bibinfo {author} {\bibfnamefont {D.}~\bibnamefont
  {Segers}}, \bibinfo {author} {\bibfnamefont {L.}~\bibnamefont
  {Dorikens-Vanpraet}}, \bibinfo {author} {\bibfnamefont {M.}~\bibnamefont
  {Dorikens}}, \bibinfo {author} {\bibfnamefont {G.}~\bibnamefont {Knuyt}},
  \bibinfo {author} {\bibfnamefont {L.~M.}\ \bibnamefont {Stals}}, \ and\
  \bibinfo {author} {\bibfnamefont {P.}~\bibnamefont {Moser}},\ }\href@noop {}
  {\bibfield  {journal} {\bibinfo  {journal} {Phys. Rev. B},\ }\textbf
  {\bibinfo {volume} {27}},\ \bibinfo {pages} {5257} (\bibinfo {year}
  {1983})}\BibitemShut {NoStop}%
\bibitem [{\citenamefont {Barrett}\ and\ \citenamefont
  {Massalski}(1966)}]{Barrett}%
  \BibitemOpen
  \bibfield  {author} {\bibinfo {author} {\bibfnamefont {C.~S.}\ \bibnamefont
  {Barrett}}\ and\ \bibinfo {author} {\bibfnamefont {T.~B.}\ \bibnamefont
  {Massalski}},\ }\href@noop {} {\emph {\bibinfo {title} {The Structure of
  Metals}}}\ (\bibinfo  {publisher} {McGraw-Hill},\ \bibinfo {address} {New
  York},\ \bibinfo {year} {1966})\BibitemShut {NoStop}%
\bibitem [{\citenamefont {Els\"asser}\ \emph
  {et~al.}(1998){\natexlab{b}}\citenamefont {Els\"asser}, \citenamefont {Zhu},
  \citenamefont {Louie}, \citenamefont {Meyer}, \citenamefont {F\"ahnle},\ and\
  \citenamefont {Chan}}]{CE2}%
  \BibitemOpen
  \bibfield  {author} {\bibinfo {author} {\bibfnamefont {C.}~\bibnamefont
  {Els\"asser}}, \bibinfo {author} {\bibfnamefont {J.}~\bibnamefont {Zhu}},
  \bibinfo {author} {\bibfnamefont {S.~G.}\ \bibnamefont {Louie}}, \bibinfo
  {author} {\bibfnamefont {B.}~\bibnamefont {Meyer}}, \bibinfo {author}
  {\bibfnamefont {M.}~\bibnamefont {F\"ahnle}}, \ and\ \bibinfo {author}
  {\bibfnamefont {C.~T.}\ \bibnamefont {Chan}},\ }\href@noop {} {\bibfield
  {journal} {\bibinfo  {journal} {Journal of Physics: Condensed Matter},\
  }\textbf {\bibinfo {volume} {10}},\ \bibinfo {pages} {5113} (\bibinfo {year}
  {1998}{\natexlab{b}})}\BibitemShut {NoStop}%
\bibitem [{\citenamefont {Els\"asser}\ \emph
  {et~al.}(1998){\natexlab{c}}\citenamefont {Els\"asser}, \citenamefont
  {Krimmel}, \citenamefont {F\"ahnle}, \citenamefont {Louie},\ and\
  \citenamefont {Chan}}]{CE3}%
  \BibitemOpen
  \bibfield  {author} {\bibinfo {author} {\bibfnamefont {C.}~\bibnamefont
  {Els\"asser}}, \bibinfo {author} {\bibfnamefont {H.}~\bibnamefont {Krimmel}},
  \bibinfo {author} {\bibfnamefont {M.}~\bibnamefont {F\"ahnle}}, \bibinfo
  {author} {\bibfnamefont {S.~G.}\ \bibnamefont {Louie}}, \ and\ \bibinfo
  {author} {\bibfnamefont {C.~T.}\ \bibnamefont {Chan}},\ }\href@noop {}
  {\bibfield  {journal} {\bibinfo  {journal} {Journal of Physics: Condensed
  Matter},\ }\textbf {\bibinfo {volume} {10}},\ \bibinfo {pages} {5131}
  (\bibinfo {year} {1998}{\natexlab{c}})}\BibitemShut {NoStop}%
\bibitem [{\citenamefont {Finnis}\ \emph
  {et~al.}(1998){\natexlab{b}}\citenamefont {Finnis}, \citenamefont {Paxton},
  \citenamefont {Methfessel},\ and\ \citenamefont {van
  Schilfgaarde}}]{Finnis98a}%
  \BibitemOpen
  \bibfield  {author} {\bibinfo {author} {\bibfnamefont {M.~W.}\ \bibnamefont
  {Finnis}}, \bibinfo {author} {\bibfnamefont {A.~T.}\ \bibnamefont {Paxton}},
  \bibinfo {author} {\bibfnamefont {M.}~\bibnamefont {Methfessel}}, \ and\
  \bibinfo {author} {\bibfnamefont {M.}~\bibnamefont {van Schilfgaarde}},\
  }\href@noop {} {\bibfield  {journal} {\bibinfo  {journal} {Phys. Rev.
  Lett.},\ }\textbf {\bibinfo {volume} {81}},\ \bibinfo {pages} {5149}
  (\bibinfo {year} {1998}{\natexlab{b}})}\BibitemShut {NoStop}%
\bibitem [{\citenamefont {Schwefel}(1977)}]{Schwefel77}%
  \BibitemOpen
  \bibfield  {author} {\bibinfo {author} {\bibfnamefont {H.-P.}\ \bibnamefont
  {Schwefel}},\ }\href@noop {} {\emph {\bibinfo {title} {{Numerische
  Optimierung von Computer--Modellen mittels der Evolutionsstrategie}}}},\
  \bibinfo {series} {Interdisciplinary Systems Research}, Vol.~\bibinfo
  {volume} {26}\ (\bibinfo  {publisher} {Birkh{\"a}user},\ \bibinfo {address}
  {Basle},\ \bibinfo {year} {1977})\BibitemShut {NoStop}%
\bibitem [{\citenamefont {Schwefel}(1993)}]{Schwefel93}%
  \BibitemOpen
  \bibfield  {author} {\bibinfo {author} {\bibfnamefont {H.-P.}\ \bibnamefont
  {Schwefel}},\ }\href@noop {} {\emph {\bibinfo {title} {Evolution and Optimum
  Seeking: The Sixth Generation}}}\ (\bibinfo  {publisher} {John Wiley},\
  \bibinfo {address} {New York},\ \bibinfo {year} {1993})\BibitemShut {NoStop}%
\bibitem [{\citenamefont {Krimmel}\ \emph
  {et~al.}(1994){\natexlab{a}}\citenamefont {Krimmel}, \citenamefont
  {Schimmele}, \citenamefont {Els{\"a}sser},\ and\ \citenamefont
  {F{\"a}hnle}}]{Krimmel94b}%
  \BibitemOpen
  \bibfield  {author} {\bibinfo {author} {\bibfnamefont {H.}~\bibnamefont
  {Krimmel}}, \bibinfo {author} {\bibfnamefont {L.}~\bibnamefont {Schimmele}},
  \bibinfo {author} {\bibfnamefont {C.}~\bibnamefont {Els{\"a}sser}}, \ and\
  \bibinfo {author} {\bibfnamefont {M.}~\bibnamefont {F{\"a}hnle}},\
  }\href@noop {} {\bibfield  {journal} {\bibinfo  {journal} {Journal of
  Physics: Condensed Matter},\ }\textbf {\bibinfo {volume} {6}},\ \bibinfo
  {pages} {7704} (\bibinfo {year} {1994}{\natexlab{a}})}\BibitemShut {NoStop}%
\bibitem [{\citenamefont {Hirth}(1980)}]{Hirth80}%
  \BibitemOpen
  \bibfield  {author} {\bibinfo {author} {\bibfnamefont {J.~P.}\ \bibnamefont
  {Hirth}},\ }\href@noop {} {\bibfield  {journal} {\bibinfo  {journal}
  {Metallurgical Transactions A},\ }\textbf {\bibinfo {volume} {11}},\ \bibinfo
  {pages} {1543} (\bibinfo {year} {1980})}\BibitemShut {NoStop}%
\bibitem [{\citenamefont {Krimmel}\ \emph
  {et~al.}(1994){\natexlab{b}}\citenamefont {Krimmel}, \citenamefont
  {Schimmele}, \citenamefont {Els{\"a}sser},\ and\ \citenamefont
  {F{\"a}hnle}}]{Krimmel94a}%
  \BibitemOpen
  \bibfield  {author} {\bibinfo {author} {\bibfnamefont {H.}~\bibnamefont
  {Krimmel}}, \bibinfo {author} {\bibfnamefont {L.}~\bibnamefont {Schimmele}},
  \bibinfo {author} {\bibfnamefont {C.}~\bibnamefont {Els{\"a}sser}}, \ and\
  \bibinfo {author} {\bibfnamefont {M.}~\bibnamefont {F{\"a}hnle}},\
  }\href@noop {} {\bibfield  {journal} {\bibinfo  {journal} {Journal of
  Physics: Condensed Matter},\ }\textbf {\bibinfo {volume} {6}},\ \bibinfo
  {pages} {7679} (\bibinfo {year} {1994}{\natexlab{b}})}\BibitemShut {NoStop}%
\bibitem [{\citenamefont {Jiang}\ and\ \citenamefont {Carter}(2004)}]{Jiang04}%
  \BibitemOpen
  \bibfield  {author} {\bibinfo {author} {\bibfnamefont {D.~E.}\ \bibnamefont
  {Jiang}}\ and\ \bibinfo {author} {\bibfnamefont {E.~A.}\ \bibnamefont
  {Carter}},\ }\href@noop {} {\bibfield  {journal} {\bibinfo  {journal} {Phys.
  Rev. B},\ }\textbf {\bibinfo {volume} {70}},\ \bibinfo {pages} {064102}
  (\bibinfo {year} {2004})}\BibitemShut {NoStop}%
\bibitem [{\citenamefont {Methfessel}\ and\ \citenamefont
  {Paxton}(1989)}]{Methfessel89}%
  \BibitemOpen
  \bibfield  {author} {\bibinfo {author} {\bibfnamefont {M.}~\bibnamefont
  {Methfessel}}\ and\ \bibinfo {author} {\bibfnamefont {A.~T.}\ \bibnamefont
  {Paxton}},\ }\href@noop {} {\bibfield  {journal} {\bibinfo  {journal} {Phys.
  Rev. B},\ }\textbf {\bibinfo {volume} {40}},\ \bibinfo {pages} {3616}
  (\bibinfo {year} {1989})}\BibitemShut {NoStop}%
\bibitem [{\citenamefont {Paxton}(2010)}]{Paxton10}%
  \BibitemOpen
  \bibfield  {author} {\bibinfo {author} {\bibfnamefont {A.~T.}\ \bibnamefont
  {Paxton}},\ }\href@noop {} {} (\bibinfo {year} {2010}),\ \bibinfo {note}
  {unpublished}\BibitemShut {NoStop}%
\bibitem [{\citenamefont {Puska}\ and\ \citenamefont
  {Nieminen}(1984)}]{Puska84}%
  \BibitemOpen
  \bibfield  {author} {\bibinfo {author} {\bibfnamefont {M.~J.}\ \bibnamefont
  {Puska}}\ and\ \bibinfo {author} {\bibfnamefont {R.~M.}\ \bibnamefont
  {Nieminen}},\ }\href@noop {} {\bibfield  {journal} {\bibinfo  {journal}
  {Phys. Rev. B},\ }\textbf {\bibinfo {volume} {29}},\ \bibinfo {pages} {5382}
  (\bibinfo {year} {1984})}\BibitemShut {NoStop}%
\bibitem [{\citenamefont {Merrill}\ and\ \citenamefont
  {Madix}(1996)}]{Merrill96}%
  \BibitemOpen
  \bibfield  {author} {\bibinfo {author} {\bibfnamefont {P.~B.}\ \bibnamefont
  {Merrill}}\ and\ \bibinfo {author} {\bibfnamefont {R.~J.}\ \bibnamefont
  {Madix}},\ }\href@noop {} {\bibfield  {journal} {\bibinfo  {journal} {Surf.
  Sci.},\ }\textbf {\bibinfo {volume} {347}},\ \bibinfo {pages} {249} (\bibinfo
  {year} {1996})}\BibitemShut {NoStop}%
\bibitem [{\citenamefont {Takai}\ \emph {et~al.}(2008)\citenamefont {Takai},
  \citenamefont {Shoda}, \citenamefont {Suzuki},\ and\ \citenamefont
  {Nagumo}}]{Takai08}%
  \BibitemOpen
  \bibfield  {author} {\bibinfo {author} {\bibfnamefont {K.}~\bibnamefont
  {Takai}}, \bibinfo {author} {\bibfnamefont {H.}~\bibnamefont {Shoda}},
  \bibinfo {author} {\bibfnamefont {H.}~\bibnamefont {Suzuki}}, \ and\ \bibinfo
  {author} {\bibfnamefont {M.}~\bibnamefont {Nagumo}},\ }\href@noop {}
  {\bibfield  {journal} {\bibinfo  {journal} {Acta Materialia},\ }\textbf
  {\bibinfo {volume} {56}},\ \bibinfo {pages} {5158} (\bibinfo {year}
  {2008})}\BibitemShut {NoStop}%
\bibitem [{\citenamefont {Kirchheim}(2010)}]{Kirchheim10}%
  \BibitemOpen
  \bibfield  {author} {\bibinfo {author} {\bibfnamefont {R.}~\bibnamefont
  {Kirchheim}},\ }\href@noop {} {\bibfield  {journal} {\bibinfo  {journal}
  {Scripta Materialia},\ }\textbf {\bibinfo {volume} {62}},\ \bibinfo {pages}
  {67} (\bibinfo {year} {2010})}\BibitemShut {NoStop}%
\bibitem [{\citenamefont {Iwamoto}\ and\ \citenamefont
  {Fukai}(1999)}]{Iwamoto99}%
  \BibitemOpen
  \bibfield  {author} {\bibinfo {author} {\bibfnamefont {M.}~\bibnamefont
  {Iwamoto}}\ and\ \bibinfo {author} {\bibfnamefont {Y.}~\bibnamefont
  {Fukai}},\ }\href@noop {} {\bibfield  {journal} {\bibinfo  {journal}
  {Materials Transactions, The Japan Inst. Metals (JIM)},\ }\textbf {\bibinfo
  {volume} {40}},\ \bibinfo {pages} {606} (\bibinfo {year} {1999})}\BibitemShut
  {NoStop}%
\bibitem [{\citenamefont {Fukai}\ \emph {et~al.}(2003)\citenamefont {Fukai},
  \citenamefont {Mori},\ and\ \citenamefont {Shinomiya}}]{Fukai03}%
  \BibitemOpen
  \bibfield  {author} {\bibinfo {author} {\bibfnamefont {Y.}~\bibnamefont
  {Fukai}}, \bibinfo {author} {\bibfnamefont {K.}~\bibnamefont {Mori}}, \ and\
  \bibinfo {author} {\bibfnamefont {H.}~\bibnamefont {Shinomiya}},\ }\href@noop
  {} {\bibfield  {journal} {\bibinfo  {journal} {J. Alloys and Compounds},\
  }\textbf {\bibinfo {volume} {348}},\ \bibinfo {pages} {105} (\bibinfo {year}
  {2003})}\BibitemShut {NoStop}%
\bibitem [{\citenamefont {Kirchheim}(2007)}]{Kirchheim07b}%
  \BibitemOpen
  \bibfield  {author} {\bibinfo {author} {\bibfnamefont {R.}~\bibnamefont
  {Kirchheim}},\ }\href@noop {} {\bibfield  {journal} {\bibinfo  {journal}
  {Acta Materialia},\ }\textbf {\bibinfo {volume} {55}},\ \bibinfo {pages}
  {5139} (\bibinfo {year} {2007})}\BibitemShut {NoStop}%
\bibitem [{Foo()}]{Footnote}%
  \BibitemOpen
  \href@noop {} {}\bibinfo {note} {This is the quantity denoted
  $E_{\text{trap}}(1,n)$ in ref~[\onlinecite{Tateyama03}] and plotted in their
  fig.~3.}\BibitemShut {Stop}%
\bibitem [{\citenamefont {Myers}\ \emph {et~al.}(1979)\citenamefont {Myers},
  \citenamefont {Picraux},\ and\ \citenamefont {Stoltz}}]{Myers79}%
  \BibitemOpen
  \bibfield  {author} {\bibinfo {author} {\bibfnamefont {S.~M.}\ \bibnamefont
  {Myers}}, \bibinfo {author} {\bibfnamefont {S.~T.}\ \bibnamefont {Picraux}},
  \ and\ \bibinfo {author} {\bibfnamefont {R.~E.}\ \bibnamefont {Stoltz}},\
  }\href@noop {} {\bibfield  {journal} {\bibinfo  {journal} {J. Appl. Phys.},\
  }\textbf {\bibinfo {volume} {50}},\ \bibinfo {pages} {5710} (\bibinfo {year}
  {1979})}\BibitemShut {NoStop}%
\bibitem [{\citenamefont {N\o{}rskov}\ \emph {et~al.}(1982)\citenamefont
  {N\o{}rskov}, \citenamefont {Besenbacher}, \citenamefont {B\o{}ttiger},
  \citenamefont {Nielsen},\ and\ \citenamefont {Pisarev}}]{Norskov82}%
  \BibitemOpen
  \bibfield  {author} {\bibinfo {author} {\bibfnamefont {J.~K.}\ \bibnamefont
  {N\o{}rskov}}, \bibinfo {author} {\bibfnamefont {F.}~\bibnamefont
  {Besenbacher}}, \bibinfo {author} {\bibfnamefont {J.}~\bibnamefont
  {B\o{}ttiger}}, \bibinfo {author} {\bibfnamefont {B.~B.}\ \bibnamefont
  {Nielsen}}, \ and\ \bibinfo {author} {\bibfnamefont {A.~A.}\ \bibnamefont
  {Pisarev}},\ }\href@noop {} {\bibfield  {journal} {\bibinfo  {journal} {Phys.
  Rev. Lett.},\ }\textbf {\bibinfo {volume} {49}},\ \bibinfo {pages} {1420}
  (\bibinfo {year} {1982})}\BibitemShut {NoStop}%
\bibitem [{\citenamefont {\ifmmode \check{C}\else
  \v{C}\fi{}\'\i{}\ifmmode~\check{z}\else \v{z}\fi{}ek}\ \emph
  {et~al.}(2004)\citenamefont {\ifmmode \check{C}\else
  \v{C}\fi{}\'\i{}\ifmmode~\check{z}\else \v{z}\fi{}ek}, \citenamefont
  {Proch\'azka}, \citenamefont {Be\ifmmode \check{c}\else
  \v{c}\fi{}v\'a\ifmmode~\check{r}\else \v{r}\fi{}}, \citenamefont
  {Ku\ifmmode~\check{z}\else \v{z}\fi{}el}, \citenamefont {Cieslar},
  \citenamefont {Brauer}, \citenamefont {Anwand}, \citenamefont {Kirchheim},\
  and\ \citenamefont {Pundt}}]{Cizek04}%
  \BibitemOpen
  \bibfield  {author} {\bibinfo {author} {\bibfnamefont {J.}~\bibnamefont
  {\ifmmode \check{C}\else \v{C}\fi{}\'\i{}\ifmmode~\check{z}\else
  \v{z}\fi{}ek}}, \bibinfo {author} {\bibfnamefont {I.}~\bibnamefont
  {Proch\'azka}}, \bibinfo {author} {\bibfnamefont {F.}~\bibnamefont
  {Be\ifmmode \check{c}\else \v{c}\fi{}v\'a\ifmmode~\check{r}\else
  \v{r}\fi{}}}, \bibinfo {author} {\bibfnamefont {R.}~\bibnamefont
  {Ku\ifmmode~\check{z}\else \v{z}\fi{}el}}, \bibinfo {author} {\bibfnamefont
  {M.}~\bibnamefont {Cieslar}}, \bibinfo {author} {\bibfnamefont
  {G.}~\bibnamefont {Brauer}}, \bibinfo {author} {\bibfnamefont
  {W.}~\bibnamefont {Anwand}}, \bibinfo {author} {\bibfnamefont
  {R.}~\bibnamefont {Kirchheim}}, \ and\ \bibinfo {author} {\bibfnamefont
  {A.}~\bibnamefont {Pundt}},\ }\href@noop {} {\bibfield  {journal} {\bibinfo
  {journal} {Phys. Rev. B},\ }\textbf {\bibinfo {volume} {69}},\ \bibinfo
  {pages} {224106} (\bibinfo {year} {2004})}\BibitemShut {NoStop}%
\bibitem [{\citenamefont {Heggie}\ \emph {et~al.}(1993)\citenamefont {Heggie},
  \citenamefont {Jones},\ and\ \citenamefont {Umerski}}]{Heggie93}%
  \BibitemOpen
  \bibfield  {author} {\bibinfo {author} {\bibfnamefont {M.~I.}\ \bibnamefont
  {Heggie}}, \bibinfo {author} {\bibfnamefont {R.}~\bibnamefont {Jones}}, \
  and\ \bibinfo {author} {\bibfnamefont {A.}~\bibnamefont {Umerski}},\
  }\href@noop {} {\bibfield  {journal} {\bibinfo  {journal} {phys. stat. sol.
  (a)},\ }\textbf {\bibinfo {volume} {138}},\ \bibinfo {pages} {383} (\bibinfo
  {year} {1993})}\BibitemShut {NoStop}%
\end{thebibliography}
\end{document}